\documentclass[aps,nofootinbib,prd,eqsecnum,twocolumn,showpacs,preprintnumbers]{revtex4}
\usepackage{textcomp}
\usepackage{amsmath}
\usepackage{amssymb}
\usepackage{esint}

\usepackage{graphicx}
\usepackage{amsfonts}
\usepackage{color}
\usepackage{bm}
\usepackage{mathrsfs}
\usepackage{epstopdf}
\usepackage{url}
\usepackage{footnote}
\usepackage{dsfont}
\usepackage{ulem}

\newcommand*{\rom}[1]{\expandafter\@slowromancap\romannumeral #1@}

\newcommand{\be}{\begin{equation}}
  \newcommand{\ee}{\end{equation}}
\newcommand{\ben}{\begin{eqnarray*}}
  \newcommand{\een}{\end{eqnarray*}}
\newcommand{\bea}{\begin{eqnarray}}
  \newcommand{\eea}{\end{eqnarray}}
\newcommand{\bdm}{\begin{displaymath}}
  \newcommand{\edm}{\end{displaymath}}
\newcommand{\ba}{\begin{align}}
  \newcommand{\ea}{\end{align}}

\begin{document}

\title{Gravitational waves and electrodynamics: New perspectives}

\author{Francisco Cabral}
\email{fc30808@alunos.fc.ul.pt}
\affiliation{Instituto de Astrof\'{\i}sica e Ci\^{e}ncias do Espa\c{c}o, Faculdade de
Ci\^encias da Universidade de Lisboa, Edif\'{\i}cio C8, Campo Grande,
P-1749-016 Lisbon, Portugal.}

\author{Francisco S. N. Lobo}
\email{fslobo@fc.ul.pt}
\affiliation{Instituto de Astrof\'{\i}sica e Ci\^{e}ncias do Espa\c{c}o, Faculdade de
Ci\^encias da Universidade de Lisboa, Edif\'{\i}cio C8, Campo Grande,
P-1749-016 Lisbon, Portugal.}

\date{\today}

\begin{abstract}
 Given the recent direct measurement of gravitational waves (GWs) by the LIGO-VIRGO collaboration, the coupling between electromagnetic fields and gravity have a special relevance since it opens new perspectives for future GW detectors and also potentially provides information on the physics of highly energetic GW sources. We explore such couplings using the field equations of electrodynamics on (pseudo) Riemann manifolds and apply it to the background of a GW, seen as a linear perturbation of Minkowski geometry. Electric and magnetic oscillations are induced that propagate as electromagnetic waves and contain information about the GW which generates them. The most relevant results are the presence of longitudinal modes and dynamical polarization patterns of electromagnetic radiation induced by GWs. These effects might be amplified using appropriate resonators, effectively improving the signal to noise ratio around a specific frequency. We also briefly address the generation of charge density fluctuations induced by GWs and the implications for astrophysics. 
\end{abstract}



\maketitle




\section{Introduction}

A century after General Relativity (GR) we have celebrated the first direct measurement of gravitational waves (GWs) by the LIGO-VIRGO collaboration \cite{Abbott:2016blz}, and ESA's LISA-Pathfinder \cite{e-LISA} science mission officially started on March, 2016. For excellent reviews on GWs see \cite{Colpi:2016fup,Sathyaprakash:2009xs}. The waves that were measured by two detectors independently, beautifully match the expected signal following a black hole binary merger, allowing the estimation of physical and kinematic properties of these black holes. This is the expected celebration which not only confirms the existence of these waves and reinforces  Einstein's GR theory of gravity, but it also marks the very birth of GW astronomy. Simultaneously, if the general relativistic interpretation of the data is correct it gives an indirect observation of black holes and the dynamics of black hole merging in binaries. 

It should be said however that GW emission from the coalescence of highly compact sources provides a test for astrophysical phenomena in the very strong gravity regime which means that a fascinating opportunity arises to study not only GR but also extended theories of gravity both classically \cite{Bogdanos:2009tn} as well as those including ``quantum corrections'' from quantum field theory  which can predict a GW signature for the non-classical physics happening at or near the black hole's horizon (see \cite{Cardoso:2016oxy,Abedi:2016hgu}  for recent claims on the detection of GW echoes in the late-time signals detected by LIGO, which point to physics beyond GR). The window is opened for the understanding of the physical nature of the sources but the consensus on the discovery of GWs is general. At the moment of the writing of this paper, after the first detection, the LIGO team has announced other two detections also interpreted as black hole binary merger events. The celebrated measurement of GW emission was done using laser interferometry, but other methods such as pulsar timing arrays \cite{Zhu:2015ara} will most probably provide positive detections in the near future. 

However, it is crucial to keep investigating different routes towards GW measurements (see \cite{Zhu:2015ara,Barausse:2014tra,Hacyan:2015kra,GWsRiles:2012yw,Singer:2014qca}) and one such route lies at the very heart of this work. Instead of using test masses and measuring the minute changes of their relative distances, as it is done in Laser interferometry (used in LIGO, VIRGO, GEO600, TAMA300 and will be used in KAGRA, LIGOIndia and LISA), we can also explore the effects of GWs on electromagnetic fields. For this purpose, one needs to compute the electromagnetic field equations on the spacetime background of a GW perturbation. This might not only provide models and simulations which can test the viability of such GW-electromagnetic detectors, but it might also contribute to a deeper understanding of the physical properties of astrophysical and cosmological sources of GWs, since these waves interact with the electromagnetic fields and plasmas which are expected to be common in many highly energetic GW sources (see \citep{Brodin:2000du}). Thus, an essential aim, in this work, will be to carefully explore the effects of GWs on electromagnetic fields.

Before approaching the issue of GW effects on electromagnetic fields, let us mention very briefly other possible routes in the quest for GW measurements. Recall that linearized gravity is also the context in which gravitoelectric and gravitomagnetic fields can be defined \citep{GravitomagnetismMashhoon:2003ax}. In particular, gravitomagnetism is associated to spacetime metrics with time-space components. Similarly, as we will see, the ($\times$) polarization of GWs is related to space-space off-diagonal metric components, which therefore resemble gravitomagnetism. This analogy might provide a motivation to explore the dynamical effects of GWs on gyroscopes. In fact, an analogy with gravitomagnetism  brings interesting perspectives regarding physical interpretations, since other analogies, in this case with electromagnetism, can be explored. In particular, gravitomagnetic effects on gyroscopes are known to be fully analogous to magnetic effects on dipoles. Now, in the case of gravitational waves these analogous (off-diagonal) effects on gyroscopes will, in general, be time dependent. 

The tiny gravitomagnetic effect on gyroscopes due to Earth's rotation, was successfully measured during the Gravity Probe B experiment \cite{Everitt:2011hp}, where the extremely small geodetic and Lens-Thirring (gravitomagnetic) deviations of the gyro's axis were measured with the help of Super Conducting Quantum Interference Devices (SQUIDS). Analogous (time varying) effects on gyroscopes due to the passage of GWs, might be measured with SQUIDS. On the other hand, rotating superconducting matter seems to generate anomalous (stronger) gravitomagnetic fields (anomalous gravitomagnetic London moment) \cite{Tajmar:2004ww,Tajmar:2006gh} so, if these results are robustly confirmed then superconductivity and super fluidity might somehow amplify gravitational phenomena. This hypothesis deserves further theoretical and experimental research as it could contribute for future advanced GW detectors.

Another promising route comes from the study of the coupling between electromagnetic fields and gravity, the topic of our concern in the present work.
Are there measurable effects on electric and magnetic fields during the passage of a GW? Could these be used in practice to study the physics of GW production from astrophysical sources, or applied to GW detection? Although very important work has been done in the past (see for example \cite{Marklund:1999sp,Brodin:2000du}), it seems reasonable to say that these routes are far from being fully explored.  

Regarding electromagnetic radiation, there are some studies regarding the effects of GWs (see for example \cite{Montanari:1998gd,Rakhmanov:2009zz}) it has been shown that gravitational waves have an important effect on the polarization of light \cite{Hacyan:2015kra}. On the other hand, lensing has been gradually more and more relevant in observational astrophysics and cosmology and it seems undoubtedly relevant to study the effects of GWs (from different types of sources) on lensing, since a GW should in principle distort any lensed image. Could lensing provide a natural amplification of the gravitational perturbation signal due to the coupling between gravity and light? These topics need careful analysis for a better understanding of the possible routes (within the reach of present technology) for gravity wave astronomy and its applications to astrophysics and cosmology.

This paper is outlined in the following manner: In Section \ref{secII}, we briefly review the foundations of electrodynamics and spacetime geometry and present the basic equations that will be used throughout this work. In Section \ref{secIII}, 
we explore the coupling between electromagnetic fields and gravitational waves.
In Section \ref{secIV}, we discuss our results and conclude.

\section{Electrodynamics: General formalism}\label{secII}

In this section, for self-consistency and self-completeness, we briefly present the general formalism that will be applied throughout the analysis. We refer the reader to \cite{FCFL}-\citep{Hehl:1999bt} for detailed descriptions of the deep relation between electromagnetic fields and spacetime geometry. We will be using a $(+---)$ signature.

Recall that in the pre-metric formalism of electrodynamics, the charge conservation provides the inhomogeneous field equations while the magnetic flux conservation is given in the homogeneous equations  \cite{HehlYuribook,Gronwald:2005tv,Hehl:2000pe,Hehl:2005hu,Hehl:1999bt}. The field equations are then given by
\begin{eqnarray}
d\bold{F}=0,  \qquad  d\bold{G}=\bold{J}.\label{fieldeqsForms} 
\end{eqnarray}
Note that these are general, coordinate free, covariant equations and there is no need for an affine or metric structure of spacetime. $\bold{J}$ is the charge current density 3-form; $\bold{G}$ is the 2-form representing the electromagnetic excitation, $\bold{F}$ is the Faraday 2-form, so that $\bold{F}=d\bold{A}$, where $\bold{A}$ is the electromagnetic potential 1-form; the operator $d$ stands for exterior derivative.

The constitutive relations (usually assumed to be linear, local, homogeneous and isotropic) between $\bold{G}$ and $\bold{F}$,
\begin{equation}
\bold{G}\longleftrightarrow\star\bold{F},\label{constitutiveForms}
\end{equation}
provide the metric structure via the Hodge star operator $\star$ , which introduce the conformal part of the metric and maps $k$-forms to $(n-k)$-forms, with $n$ the dimension of the spacetime manifold. On these foundations, the electromagnetic field equations on the background of a general (pseudo) Riemann spacetime manifold can be obtained. In the tensor formalism we get 
\begin{equation}
\partial_{\mu}F^{\mu\nu}+\dfrac{1}{\sqrt{-g}}\partial_{\mu}(\sqrt{-g})F^{\mu\nu}=\mu_{0}j^{\nu},\quad \partial_{[\alpha}F_{\beta\gamma]}=0,\label{Maxwelleqstensor}
\end{equation}
where we have used in the inhomogeneous equations the general expression for the divergence of anti-symmetric tensors in pseudo-Riemann geometry, $\nabla_{\mu}\Theta^{\mu\nu}=\dfrac{1}{\sqrt{-g}}\partial_{\mu}\left( \sqrt{-g}\Theta^{\mu\nu}\right)$. The homogeneous equations are independent from the metric (and connection) due to the torsionless character of Riemann geometry.

We introduce the definitions 
\begin{eqnarray}
F_{0k}&=&\dfrac{1}{c}\partial_{t}A_{k}-\partial_{k}A_{0}\equiv \dfrac{E_{k}}{c},
   \\
F_{jk}&=&\partial_{j}A_{k}-\partial_{k}A_{j}\equiv -\epsilon_{ijk}B^{i}
,\label{magneticfield}
\end{eqnarray}
where $\epsilon_{ijk}$ is the totally antisymmetric 3-dimensional Levi-Civita (pseudo) tensor, $B^{i}$ is a vector density (a natural surface integrand) and $E_{k}$ is a co-vector (a natural line integrand). Then, the homogeneous equations are the usual Faraday and magnetic Gauss laws
\begin{equation}
\partial_{t}B^{i}+\epsilon^{ijk}\partial_{j}E_{k}=0, \qquad  \partial_{j}B^{j}=0,
\end{equation}
while the inhomogeneous equations can be separated into the generalized Gauss and Maxwell-Amp\`{e}re laws. These are, respectively
\begin{eqnarray}
\partial_{k}E_{j}\left(g^{0k}g^{j0}-g^{jk}g^{00}\right)
-\partial_{\mu}B^{k}cg^{m\mu}g^{n0}\epsilon_{kmn} \nonumber \\
+E_{j}\gamma^{j}
-B^{k}c\epsilon_{kmn}\sigma^{mn0}
=\dfrac{\rho}{\varepsilon_{0}},
   \label{GaussGenerall}
\end{eqnarray}
\begin{eqnarray}
\dfrac{1}{c}\partial_{\mu}E_{j}\left(g^{0\mu}g^{ji}-g^{j\mu}g^{0i}\right)
-\epsilon_{kmn}\partial_{\mu}B^{k}g^{m\mu}g^{ni}  \nonumber \\
+E_{j}\xi^{ji}
-B^{k}\epsilon_{kmn}\sigma^{mni}=\mu_{0}j^{i}, \label{MaxAmpereGenerall}
\end{eqnarray}
where
\begin{eqnarray}
\gamma^{j} &\equiv & \Big[\partial_{k}\left(g^{0k}g^{j0}-g^{jk}g^{00}\right)
    \nonumber \\
&&+\dfrac{1}{\sqrt{-g}}\partial_{k}(\sqrt{-g})\left(g^{0k}g^{j0}-g^{jk}g^{00}\right)\Big],
\end{eqnarray}
\begin{equation}
\sigma^{mn\beta}\equiv \left[\partial_{\mu}(g^{m\mu}g^{n\beta})+\dfrac{1}{\sqrt{-g}}\partial_{\mu}(\sqrt{-g})(g^{m\mu}g^{n\beta}) \right],
\end{equation}
\begin{eqnarray}
\xi^{ji}&\equiv &\dfrac{1}{c}\Big[\partial_{\mu}\left(g^{0\mu}g^{ji}-g^{j\mu}g^{0i}\right)
   \nonumber \\
&&+\dfrac{1}{\sqrt{-g}}\partial_{\mu}(\sqrt{-g})\left(g^{0\mu}g^{ji}-g^{j\mu}g^{0i}\right)\Big] \,.
\end{eqnarray}

One sees clearly, that new electromagnetic phenomena are expected due to the presence of extra electromagnetic couplings induced by spacetime curvature. In particular, the magnetic terms in the Gauss law are only present for non-vanishing off-diagonal time-space components $g^{0j}$, which in linearized gravity correspond to the gravitomagnetic potentials. These terms are typical of axially symmetric geometries (see \cite{FCFLapplications,Rezzolla:2000dk}).

For diagonal metrics, the inhomogeneous equations, the Gauss and Maxwell Amp\`{e}re laws, can be recast into the following forms
\begin{equation}
-g^{kk}g^{00}\partial_{k}E_{k}+E_{k}\gamma^{k}=\dfrac{\rho}{\varepsilon_{0}},\label{gausssimple}
\end{equation}
\begin{equation}
\epsilon_{ijk}g^{ii}g^{jj}\partial_{j}B^{k}+\dfrac{1}{c^{2}}g^{00}g^{ii}\partial_{t}E_{i}+\epsilon_{ijk}\sigma^{ji
i}B^{k}+E_{i}\xi^{ii}=\mu_{0}j^{i},\label{MaxAmp2}
\end{equation}
with
\begin{equation}
\gamma^{k}(\bold{x})\equiv\ -\left[g^{kk}g^{00}\dfrac{1}{\sqrt{-g}}\partial_{k}(\sqrt{-g})+\partial_{k}\left(g^{kk}g^{00}\right)\right],\label{funcgauss}
\end{equation}
and
\begin{eqnarray}
\sigma^{jii}(\bold{x}) \equiv  g^{jj}g^{ii}\dfrac{1}{\sqrt{-g}}\partial_{j}(\sqrt{-g})+\partial_{j}(g^{jj}g^{ii}),\label{sigmaximaxamp}
\end{eqnarray}
\begin{equation}
\xi^{ii}(\bold{x}) \equiv  g^{00}g^{ii}\dfrac{1}{c^{2}}\dfrac{1}{\sqrt{-g}}\partial_{t}(\sqrt{-g})+\dfrac{1}{c^{2}}\partial_{t}(g^{00}g^{ii}).\label{ximaxamp}
\end{equation}
The Einstein summation convention is applied in Eq. (\ref{MaxAmp2}) only for $j$ and $k$ while the index $i$ is fixed by the right hand side. Also, no contraction is assumed in Eq. (\ref{funcgauss}) nor in the expression for $\sigma^{jii}$.

New electromagnetic effects induced by the spacetime geometry include an inevitable spatial variability (non-uniformity) of electric fields whenever we have non-vanishing geometric functions $\gamma^{k}$, electromagnetic oscillations (therefore waves) induced by gravitational radiation and also additional electric contributions to Maxwell's displacement current in the generalized Maxwell-Amp\`{e}re law. This last example becomes clearer by re-writing Eq. (\ref{MaxAmp2}) in the form
\begin{equation}
\epsilon_{ijk}\partial_{j}\bar{B}^{iijjk}=\mu_{0}(\jmath^{i}+\jmath^{i}_{D}),\label{maxampdisplacement}
\end{equation}
where $\jmath^{i}\equiv \sqrt{-g}j^{i}$ and
\begin{eqnarray}
\jmath^{i}_{D}&\equiv & -\varepsilon_{0}\sqrt{-g}\left(g^{00}g^{ii}\partial_{t}E_{i}+c^{2}E_{i}\xi^{ii}\right), \label{maxdisplacement}
\nonumber  \\
& =&-\varepsilon_{0}\partial_{t}(\sqrt{-g}g^{00}g^{ii}E_{i})
\end{eqnarray}
is the generalized Maxwell displacement current density and
\begin{equation}
\bar{B}^{iijjk}\equiv g^{ii}g^{jj}\sqrt{-g}B^{k}. 
\end{equation}

Again no summation convention is assumed for the index $i$ in Eqs. (\ref{maxampdisplacement}) and (\ref{maxdisplacement}). The functions $\xi^{ii}$ vanish for stationary spacetimes but might have an important contribution for strongly varying gravitational waves (high frequencies), since they depend on the time derivatives of the metric. 
Analogously, Eq. (\ref{gausssimple}) can be written as
\begin{equation}
\partial_{k}\tilde{E}^{k}=\dfrac{\varrho}{\varepsilon_{0}},\label{gausschangevar}
\end{equation}
where
\begin{equation}
\tilde{E}^{j}\equiv -g^{jj}g^{00}\sqrt{-g}E_{j}, \qquad \varrho\equiv \sqrt{-g}\rho.
\label{gausschangevar2}
\end{equation}

These are physical, observable effects of spacetime geometry in electromagnetic fields expressed in terms of the extended Gauss and Maxwell-Amp\`{e}re laws which help the comparison with the usual inhomogeneous equations in Minkowski spacetime, making clearer the physical interpretations of such effects. 

Finally, we review the field equations in terms of the electromagnetic 4-potential which in vacuum are convenient to study electromagnetic wave phenomena.
From the definition of the Faraday tensor and Eq. (\ref{Maxwelleqstensor}), we get
\begin{equation}
\nabla_{\mu}\nabla^{\mu}A^{\nu}-g^{\lambda\nu}R_{\varepsilon\lambda}A^{\varepsilon}-\nabla^{\nu}\left( \nabla_{\mu}A^{\mu}\right)=\mu_{0}j^{\nu},  
\end{equation}
where $R_{\varepsilon\lambda}$ is the Ricci tensor. Using the expression for the (generalized) Laplacian in pseudo-Riemann manifolds, $\nabla_{\mu}\nabla^{\mu}\psi=
\dfrac{1}{\sqrt{-g}}\partial_{\mu}\left( \sqrt{-g}g^{\mu\lambda}\partial_{\lambda}\psi\right)$, and assuming the Generalized Lorenz gauge ($\nabla_{\mu}A^{\mu}=0$) in vacuum, we arrive at
\begin{equation}
\partial_{\mu}\partial^{\mu}A^{\nu}+
\dfrac{1}{\sqrt{-g}}\partial_{\mu}\left( \sqrt{-g}g^{\mu\lambda}\right)\partial_{\lambda}A^{\nu}-g^{\lambda\nu}R_{\varepsilon\lambda}A^{\varepsilon}=0.\label{Proca}   
\end{equation}
For a diagonal metric, we get
\begin{equation}
\partial_{\mu}\partial^{\mu}A^{\nu}+
\dfrac{1}{\sqrt{-g}}\partial_{\mu}\left( \sqrt{-g}g^{\mu\mu}\right)\partial_{\mu}A^{\nu}-g^{\nu\nu}R_{\varepsilon\nu}A^{\varepsilon}=0,\label{wavepotential} 
\end{equation}
with no contraction assumed in $\nu$. In general, and contrary to electromagnetism in Minkowski spacetime, the equations for the components of the electromagnetic 4-potential are coupled even in the (generalized) Lorenz gauge. Notice also that for Ricci-flat spacetimes, the term containing the Ricci tensor vanishes. Naturally, the vacuum solutions of GR are examples of such cases. New electromagnetic phenomena are expected to be measurable, for gravitational fields where the geometric dependent terms in Eq. (\ref{Proca}) are significant.

This completes the main axiomatic (foundational) formalism of electrodynamics in the background of curved (pseudo-Riemann) spacetime.

\section{Gravitational waves and electromagnetic fields}\label{secIII}

As mentioned in the Introduction, GWs have been recently detected by the LIGO team using laser interferometry \cite{Abbott:2016blz}. Another method that has been carried over the last decade to detect GWs is that of pulsar timing arrays. Nevertheless, it is crucial to keep exploring different routes towards GW detection and its applications to astrophysics and cosmology.
Due to the huge distances in the Cosmos, any GW reaching Earth should have an extremely low amplitude. Therefore, the linearisation of gravity is usually applied which allows to derive the wave equations. It is a perturbative approach which is background dependent and its common to consider a Minkowski background. In any case, the GW can be seen as a manifestation of propagating spacetime geometry perturbations. 

In principle, the passage of a GW in a region with electromagnetic fields will have a measurable effect. To compute this we have to consider Maxwell's equations on the perturbed background of a GW. We shall consider a GW as a perturbation of Minkowski spacetime given by $g_{\alpha\beta}= \eta_{\alpha\beta}+h_{\alpha\beta}$, with $\vert h_{\alpha\beta}\vert\ll 1$, so that
\begin{equation}
ds^{2}=c^{2}dt^{2}-dx^{2}-dy^{2}-dz^{2}+h_{\alpha\beta}dx^{\alpha}dx^{\beta},
\end{equation}
where the perturbation corresponds to a wave travelling along the $z$ axis, i.e.,
\begin{eqnarray}
ds^{2}&=&c^{2}dt^{2}  -dz^{2}-[1-f_{+}(z-ct)]dx^{2}
    \nonumber \\
&-&[1+f_{+}(z-ct)]dy^{2}+2f_{\times}(z-ct)dxdy , \label{gwmetric}
\end{eqnarray}
and $(+)$ and $(\times)$ refer to the two independent polarizations characteristic of GWs in GR. This metric is a solution of Einstein's field equations in the linear approximation, in the so-called TT (Transverse-Traceless) Lorenz Gauge.
For this metric, we get
\begin{equation}   
\dfrac{1}{\sqrt{-g}}\partial_{z}(\sqrt{-g})=\dfrac{f_{\times}\left(\partial_{z}f_{\times} \right)+f_{+}\left(\partial_{z}f_{+} \right) }{f_{\times}^{2}+f_{+}^{2}-1},\label{derivzdet}
\end{equation}
\begin{equation}
\dfrac{1}{\sqrt{-g}}\partial_{t}(\sqrt{-g})=\dfrac{f_{\times}\left(\partial_{t}f_{\times} \right)+f_{+}\left(\partial_{t}f_{+} \right) }{f_{\times}^{2}+f_{+}^{2}-1}\label{derivtdet}.
\end{equation}
These quantities will be useful further on.

\subsection{GW effects in electric and magnetic fields}

Consider an electric field in the background of a GW travelling in the $z$ direction. The general expression for Gauss' law (\ref{GaussGenerall}), in vacuum, is now given by 
\begin{eqnarray}
[1-f_{+}(z,t)]^{-1}\partial_{x}E_{x}+
[1+f_{+}(z,t)]^{-1}\partial_{y}E_{y}
   \nonumber  \\
+\partial_{z}E_{z} -f_{\times}^{-1}(z,t)\left(\partial_{y}E_{x}+\partial_{x}E_{y} \right)
  \nonumber  \\ 
+\left[ 
\dfrac{1}{\sqrt{-g}}\partial_{z}(\sqrt{-g}) \right]E_{z} = 0,\label{gwgaussgeral}
\end{eqnarray}
which clearly shows that physical (possibly observable) effects are induced by the propagation of GWs.

As for the Maxwell-Amp\`{e}re law, Eq. (\ref{MaxAmpereGenerall}) provides the following relations in vacuum:
\begin{eqnarray}
&&\frac{1}{c^{2}}\left[f_{\times}^{-1}\partial_{t}E_{y}
-(1-f_{+})^{-1}\partial_{t}E_{x}\right]+E_{x}\xi^{xx}+E_{y}\xi^{yx}
   \nonumber  \\
&&-(1-f_{+})^{-1}\left[(1+f_{+})^{-1}\partial_{y}B^{z}-\partial_{z}B^{y}\right]-B^{y}\sigma^{zxx}
   \nonumber  \\
&&+B^{x}\sigma^{zyx}-f_{\times}^{-1}\left(f_{\times}^{-1}\partial_{y}B^{z}+\partial_{z}B^{x}\right)
=0 \,,
\end{eqnarray}
\begin{eqnarray}
&&\dfrac{1}{c^{2}}\left[f_{\times}^{-1}
\partial_{t}E_{x}-(1+f_{+})^{-1}\partial_{t}E_{y}\right]+E_{y}\xi^{yy}
+E_{x}\xi^{xy}
   \nonumber   \\
&&-(1+f_{+})^{-1}\left[(1-f_{+})^{-1}\partial_{x}B^{z}-\partial_{z}B^{x}\right] +B^{x}\sigma^{zyy}
   \nonumber   \\
&&-B^{y}\sigma^{zxy}
+f_{\times}^{-1}\left(f_{\times}^{-1}\partial_{x}B^{z}+\partial_{z}B^{y}\right)
=0 \,,
\end{eqnarray}
\begin{eqnarray}
&&-\dfrac{1}{c^{2}}\partial_{t}E_{z}+E_{z}\xi^{zz}
-f_{\times}^{-1}\left( \partial_{y}B^{y}-
\partial_{x}B^{x}\right) 
    \nonumber   \\
&&
+\left[(1-f_{+})^{-1}\partial_{x}B^{y}-
(1+f_{+})^{-1}\partial_{y}B^{x}\right] 
=0 \,,
\label{Gwmaxamp}
\end{eqnarray}
with the non-vanishing geometric coefficients given by
\begin{equation}
\xi^{xx}=\dfrac{1}{c^{2}}\dfrac{f_{\times}(f_{+}-1)\partial_{t}f_{\times}-(f_{\times}^{2}+f_{+}-1)\partial_{t}f_{+}}{(f_{+}-1)^{2}(f_{\times}^{2}+f_{+}^{2}-1)},
 \nonumber
\end{equation}
\begin{equation}
\xi^{yx}=\xi^{xy}=\dfrac{1}{c^{2}}\dfrac{-(f_{+}^{2}-1)\partial_{t}f_{\times}+f_{\times}f_{+}\partial_{t}f_{+}}{f_{\times}^{2}(f_{\times}^{2}+f_{+}^{2}-1)},
 \nonumber
\end{equation}
%
%
\begin{equation}
\xi^{yy}=\dfrac{1}{c^{2}}\dfrac{-f_{\times}(f_{+}+1)\partial_{t}f_{\times}+(f_{\times}^{2}-f_{+}-1)\partial_{t}f_{+}}{(f_{+}+1)^{2}(f_{\times}^{2}+f_{+}^{2}-1)},
 \nonumber
\end{equation}
\begin{equation}
\xi^{zz}=-\dfrac{1}{c^{2}}\dfrac{f_{\times}\left(\partial_{t}f_{\times} \right)+f_{+}\left(\partial_{t}f_{+} \right) }{f_{\times}^{2}+f_{+}^{2}-1},
 \nonumber
\end{equation}
\begin{equation}
\sigma^{zxx}=-\dfrac{f_{\times}(f_{+}-1)\partial_{z}f_{\times}-(f_{\times}^{2}+f_{+}-1)\partial_{z}f_{+}}{(f_{+}-1)^{2}(f_{\times}^{2}+f_{+}^{2}-1)}, 
 \nonumber
\end{equation}
\begin{equation}
\sigma^{zyy}=-\dfrac{-f_{\times}(f_{+}+1)\partial_{z}f_{\times}+(f_{\times}^{2}-f_{+}-1)\partial_{z}f_{+}}{(f_{+}+1)^{2}(f_{\times}^{2}+f_{+}^{2}-1)}, 
 \nonumber
\end{equation}
\begin{equation}
\sigma^{zxy}=\sigma^{zyx}=-\dfrac{-(f_{+}^{2}-1)\partial_{z}f_{\times}+f_{\times}f_{+}\partial_{z}f_{+}}{f_{\times}^{2}(f_{\times}^{2}+f_{+}^{2}-1)}.
 \nonumber
\end{equation}

A natural consequence of these laws is the generation of electromagnetic waves induced by gravitational radiation. Initially static electric and magnetic fields become time dependent during the passage of GWs which might be detectable in this way.

In general, the system of coupled Eqs. (\ref{GaussGenerall})-(\ref{MaxAmpereGenerall}) and the homogeneous equations in (\ref{Maxwelleqstensor})  have to be taken as a whole. As we will see from Eq. (\ref{gwgaussgeral}), in some specific situations the electric field can be solved directly from Gauss' law. This electric field can in turn act as a source for magnetism via the Maxwell-Amp\`{e}re relations in Eqs. (3.6)-(\ref{Gwmaxamp}), where the presence of the GW induces extra terms proportional to the electric field. 
In this work, we will explore relatively simple situations in order to illustrate the effects of GWs in electric and magnetic fields.
Let us start by considering the effects of GWs in electric fields.

\subsection{Electric field oscillations induced by GWs}

We will consider electric fields in the following three scenarios.

\subsubsection{Electric field aligned with the z axis}

An electric field along the $z$ axis can easily be achieved by charged plane plates constituting a capacitor. In the absence of GWs the electric field thus produced is approximately uniform (neglecting boundary effects) for static uniform charge distributions or time variable if there is an alternate current (as in the case of a RLC circuit with a variable voltage signal generator).
With the passage of the GW, in general the electric field is perturbed by both the ($+$) and ($\times$) modes. 
To see this let us look at Gauss' law when the electric field is aligned with the direction of the GW propagation. From Eq. (\ref{gwgaussgeral}), we have
\begin{equation}
\partial_{z}E_{z}+E_{z}\left[
\dfrac{1}{\sqrt{-g}}\partial_{z}(\sqrt{-g}) \right]=0,
\end{equation}
where
$   
\dfrac{1}{\sqrt{-g}}\partial_{z}(\sqrt{-g})$ is given by the expression in Eq. (\ref{derivzdet}). We can see that even if in the absence of any GW the electric field was static and uniform, during the passage of the spacetime disturbance, the field will be time varying and non-uniform, oscillating with the same frequency of the passing GW. In fact, the general solution is
\begin{equation}
E_{z}(x,y,z,t)=\dfrac{E_{0}}{\sqrt{-g}}=\dfrac{E_{0}}{\sqrt{1-f_{+}^{2}-f_{\times}^{2}}}
,\label{gwelectricz}
\end{equation}
where in the most general case, $E_{0}=E_{0}(x,y;t)$. To get the full description of the electric field one has to consider also both the Maxwell-Amp\`{e}re relations in Eqs. (3.4)-(\ref{Gwmaxamp}) and the Faraday law. Nevertheless it is already clear from Eq. (\ref{gwelectricz}) that GWs induce propagating electric oscillations.  
 
We will consider the most simple case in which $E_{0}$ is a constant (without any dependence on $x$,$y$ or $t$). Indeed, one can easily verify that the fields $\bold{E}=(0,0,E_{z})$, $\bold{B}=(0,0,0)$ constitute a (trivial) solution of the full Maxwell equations, namely Eqs. (\ref{gwgaussgeral})-(\ref{Gwmaxamp}) together with the homogeneous equations in (\ref{Maxwelleqstensor}). Notice that for zero magnetic field the $z$ Maxwell-Amp\`{e}re equation (\ref{Gwmaxamp}) is
\begin{equation}
-\dfrac{1}{c^{2}}\partial_{t}E_{z}+E_{z}\xi^{zz}=0,\label{electricfluxconserved}
\end{equation}
which is verified by the solution in (\ref{gwelectricz}) for a constant $E_{0}$. This can easily be seen when one considers that
\begin{equation}   
\xi^{zz}=-\dfrac{1}{c^{2}}
\dfrac{1}{\sqrt{-g}}\partial_{t}(\sqrt{-g}),
\end{equation}  
in accordance with the expressions previously shown for $\xi^{zz}$ and Eq. (\ref{derivtdet}). In this case, the coupling between the electric field and the GW in the expression for the generalized Maxwell displacement current density,  compensates the traditional term which depends on the time derivative of the electric field. In fact, by multiplying by $c^{2}$, then Eq. (\ref{electricfluxconserved}) can be interpreted as the conservation of the total electric flux density. This situation is thus compatible with the experimental scenario where there are no currents producing any magnetic field and the electric field, although changing in time, due to the coupling with gravity does not give rise to any magnetic field, since the total electric flux is conserved. Naturally, this is not the general case. For example in the presence of currents along the $z$ axis, $B^{x},B^{y}\neq 0$ and due to the gravitational factors in the equations (3.6)-(\ref{Gwmaxamp}) the magnetic field is dynamical (time dependent). Therefore, this field necessarily affects the electric field via the Faraday law,
\begin{equation}
\partial_{y}E_{z}=\dfrac{\partial_{y}E_{0}}{\sqrt{-g}}=-\partial_{t}B^{x},\qquad \partial_{x}E_{z}=\dfrac{\partial_{x}E_{0}}{\sqrt{-g}}=\partial_{t}B^{y}, 
\end{equation}
which implies that in general $E_{0}=E_{0}(x,y,t)$. Since $E_{0}$ is time dependent, in such a case the electric field contributes to the magnetic field via the (non-null) generalized Maxwell displacement current, in accordance with Eq. (\ref{Gwmaxamp}), where now 
\begin{equation}
-\dfrac{1}{c^{2}}\partial_{t}E_{z}+E_{z}\xi^{zz}\neq 0.
\end{equation}  
  
As a practical application consider the following harmonic GW perturbation
\begin{eqnarray}
f_{+}(z,t)&=&a\cos \left(kz-wt \right), \label{harmonicgwa}\\ 
f_{\times}(z,t)&=&b\cos \left(kz-wt+\alpha \right).\label{harmonicgw}  
\end{eqnarray}
In this case, we get the following electric oscillations
\begin{eqnarray}
E_{z}(z,t)&=&\tilde{E}_{0}\big[1-a^{2}\cos^{2}\left(kz-wt \right)
     \nonumber \\
&& -b^{2}\cos^{2} \left(kz-wt+\alpha\right)\big]^{-1/2}\label{gwelectricz2},
\end{eqnarray}
for $a^{2}+b^{2}\leq 1$, which is obeyed by the extremely low amplitude GWs reaching the Solar System. Here $\tilde{E}_{0}$ is an arbitrary fixed constant and $\alpha$ is the phase difference. These electric oscillations will show distinct features sensitive to the $(+)$ or $(\times)$ GW modes. It provides a window for detecting and analysing GW signals directly converted into electromagnetic information. 

Notice that the electric waves produced are longitudinal, since these are propagating along the same direction of the GW, even though the electric field is aligned with this direction. To grasp the physical interpretation behind this non-intuitive result, recall that the electric energy density depends quadratically on the field and therefore it is the energy density fluctuation induced by the GW which propagates along the direction of $\bold{k}=k^{z}e_{z}$. 

In order to have an approximate idea on the energy density $u^{em}$ of the resulting electromagnetic wave we can use the usual expression (derived from Maxwell electrodynamics in Minkowski spacetime). We get
\begin{eqnarray}
u^{em}\sim\varepsilon_{0}E_{z}^{2}(z,t)&=&\varepsilon_{0}\tilde{E}_{0}^{2}\big[1-a^{2}\cos^{2}\left(kz-wt \right)
     \nonumber  \\
&-&b^{2}\cos^{2} \left(kz-wt+\alpha\right) \big]^{-1} ,
\end{eqnarray} 
and the energy per unit area and unit time through any surface (with a normal making an angle $\vartheta$ with the $z$ axis) is approximately expressed by
\begin{eqnarray}
\Vert\vec{S}\Vert \cos \vartheta =\varepsilon_{0}c\tilde{E}_{0}^{2}\big[1-a^{2}\cos^{2}\left(kz-wt \right)
     \nonumber  \\
-b^{2}\cos^{2} \left(kz-wt+\alpha\right) \big]^{-1} \cos \vartheta,
\end{eqnarray}
where $\vec{S}$ is the Poynting vector, and $S\equiv u^{em}c$.

If $\tilde{E}_{0}$ is the electric field in the absence of GWs, then the relevant dimensionless quantity to be measured is given by the following expression 
\begin{eqnarray}
\Bigg\vert\dfrac{ E_{z}(z,t)-\tilde{E}_{0}}{\tilde{E}_{0}}\Bigg\vert= \big| \big[1-a^{2}\cos^{2}\left(kz-wt \right)
     \nonumber \\
-b^{2}\cos^{2} \left(kz-wt+\alpha\right)\big]^{-1/2}-1 \big|\,,
\end{eqnarray}
and in terms of the energy density, we get
\begin{eqnarray}
\Bigg\vert\dfrac{ u^{em}(z,t)-u^{em}_{0}}{u^{em}_{0}}\Bigg\vert=\big\vert\big[1-a^{2}\cos^{2}\left(kz-wt \right)
      \nonumber \\
-b^{2}\cos^{2} \left(kz-wt+\alpha\right)\big]^{-1}-1\big\vert,
\end{eqnarray}
with $u^{em}_{0}=\varepsilon_{0}\tilde{E}_{0}^{2}$.

Substituting in these two expressions the typical amplitudes for GWs due to binaries ($10^{-25}-10^{-21}$), the induced electric field and corresponding energy density oscillations signal will be extremely small. 
Concerning GWs reaching the Solar System, the detectors which might have a response proportional to the electric field magnitude or rather to its energy (proportional to the square of the electric field magnitude), must be extremely sensitive.
We emphasize the fact that, in principle, this electromagnetic wave can be confined in a cavity using very efficient reflectors for the frequency $w$. Then, under appropriate (resonant) geometric conditions, the signal can be amplified. This might have very important practical applications for future GW detectors.

\subsubsection{Electric field in the xy plane}

Suppose we have an electric field in the $x$ direction. The electric field could initially be uniform and confined within a plane capacitor. In these conditions, the Gauss law in vacuum becomes
\begin{equation}
[1-f_{+}(z,t)]^{-1}\dfrac{\partial E_{x}}{\partial x}-(f_{\times})^{-1}(z,t)\dfrac{\partial E_{x}}{\partial y}=0.\label{gwgaussxcomp}
\end{equation}
A similar expression is obtained if the electric field is aligned with the $y$ axis. Assuming a separation of variables $E_{x}(x,y,z,t)=F_{1}(x,z,t)F_{2}(y,z,t)$, where $z$ and $t$ are seen as external parameters, substituting in the above equation and dividing it by $E_{x}$, we obtain
\begin{equation}
(1-f_{+})^{-1}\dfrac{\partial_{x}F_{1}}{F_{1}}=f_{\times}^{-1}\dfrac{\partial_{y}F_{2}}{F_{2}},
\end{equation}
therefore, one arrives at the following expressions 
\begin{equation}
F_{1}(x;z)=C_{1}(z,t)e^{-\left(1-f_{+}\right)x },\quad F_{2}(x;z)=C_{2}(z,t)e^{-f_{\times}y}.
\end{equation}
Since we can always add a constant to the solution obtained from $F_{1}(x,z,t)F_{2}(y,z,t)$, we can write
\begin{equation}
E_{x}(x,y,z,t)=E_{0x}\left[1+\tilde{C}(z,t)e^{-\left[\left(1-f_{+}\right)x+f_{\times}y \right] }\right],\label{electricgwx}
\end{equation}
where in general $\tilde{C}(z,t)=\tilde{C}\left[f_{+}(z,t),f_{\times}(z,t) \right]$ can be obtained by taking into account the other Maxwell equations. The full solution should be compatible with the limit without gravity in which we recover the uniform field $E_{x}=E_{0x}$. Therefore
\begin{equation}
f_{+}=f_{\times}=0\Rightarrow \tilde{C}(z,t)=0.
\end{equation}

A natural Anszatz is
\begin{equation}
\tilde{C}(z,t)=\eta f^{\alpha_{1}}_{+}+\beta f^{\alpha_{2}}_{\times}+\mu f^{\alpha_{3}}_{+}f^{\alpha_{4}}_{\times},
\end{equation}
where $\eta,\beta,\mu$ and $\alpha_{i}$ ($i=1,2,3,4$) are constants. But as previously said the form of this function can be studied by considering compatibility with the remaining Maxwell equations.

For the harmonic GW introduced before, the second term in the solution above, Eq. (\ref{electricgwx}) is given by the following expression
\begin{eqnarray}
E_{0x}\tilde{C}(z,t)\exp\big\{-
\big[\left( 1-a\cos\left(kz-wt\right)\right)x
\nonumber  \\
+b\cos\left(kz-wt+\alpha\right)y\big]  \big\}.
\label{gwelectricx}
\end{eqnarray}

The solution obtained is also sensitive to the existence or not of two modes in the GW, to their amplitudes and phase difference.
Although this solution obeys the Gauss law, it  implies a non-zero dynamical magnetic field, according to Faraday's law. As mentioned, to get a full treatment one should then check the consistency with the other Maxwell equations, in order to derive restrictions on the mathematical form of $\tilde{C}(z,t)$. 

Let us consider now the case where an electric field $\bold{E_{1}}=(E_{x},0)$ is generated by a plane capacitor oriented along the $x$ axis and a second electric field $\bold{E_{2}}=(0,E_{y})$ is generated by another similar capacitor oriented along the $y$ axis.
In this condition, the resulting electric field in the vacuum between the charged plates, $\bold{E}=\bold{E_{1}}+\bold{E_{2}}=(E_{x},E_{y})$, obeys the equation  
\begin{eqnarray}
(1-f_{+})^{-1}\partial_{x}E_{x} &-& (f_{\times})^{-1}\partial_{y}E_{x}
+(1+f_{+})^{-1}\partial_{y}E_{y}
\nonumber \\
&&-(f_{\times})^{-1}\partial_{x}E_{y}=0.
\end{eqnarray}
A possible solution to this equation is given by
\begin{eqnarray}
E_{x}(x,y,z,t)=E_{0x}\left[1+\tilde{C_{1}}(z,t)e^{-\left[\left(1-f_{+}\right)x+f_{\times}y \right] }\right],
\nonumber \\ \label{electricfieldxyplanex}
\end{eqnarray}
\begin{eqnarray}
E_{y}(x,y,z,t)=E_{0y}\left[1+\tilde{C_{2}}(z,t)e^{-\left[f_{\times}x+\left(1+f_{+}\right)y \right] }\right],
\nonumber \\ \label{electricfieldxyplaney}
\end{eqnarray}
where for $f_{+}=f_{\times}=0$ we get $\tilde{C_{1}}(z,t)=\tilde{C_{2}}(z,t)=0$. 
\begin{figure}[h]
\centering
\includegraphics[width=7.0cm]{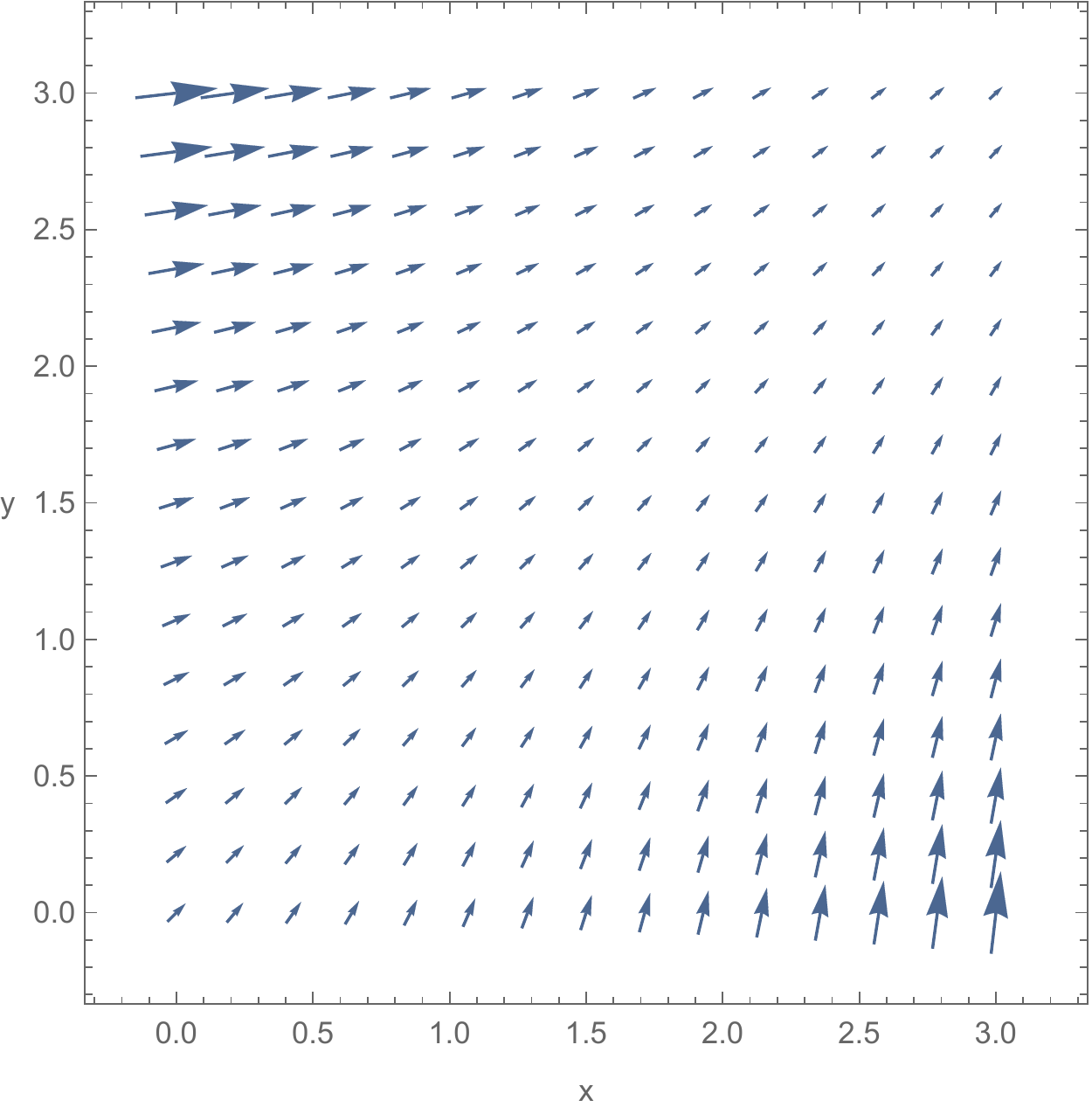} 
\hspace{1.5cm}
\caption{This vector plot illustrates the spatial (non-linear) polarization pattern on the $(x,y)$ plane for the electric field at a given instant. This pattern is exclusively induced by the GW. The GW parameters are: $a=0.036$, $b=0.766$, $w/2\pi=89.81Hz$, $\alpha=0.11\pi$. We have used Eqs. (\ref{electricfieldxyplanex}) and (\ref{electricfieldxyplaney}), where for simplicity we assumed $\tilde{C_{1}}(z,t)=\tilde{C_{2}}(z,t)=1$ and electric field magnitudes $E_{0x}=E_{0y}=10^{-3}V/m$.}
\label{fig1}
\end{figure}
\begin{figure}[h]
\centering
\includegraphics[width=7.0cm]{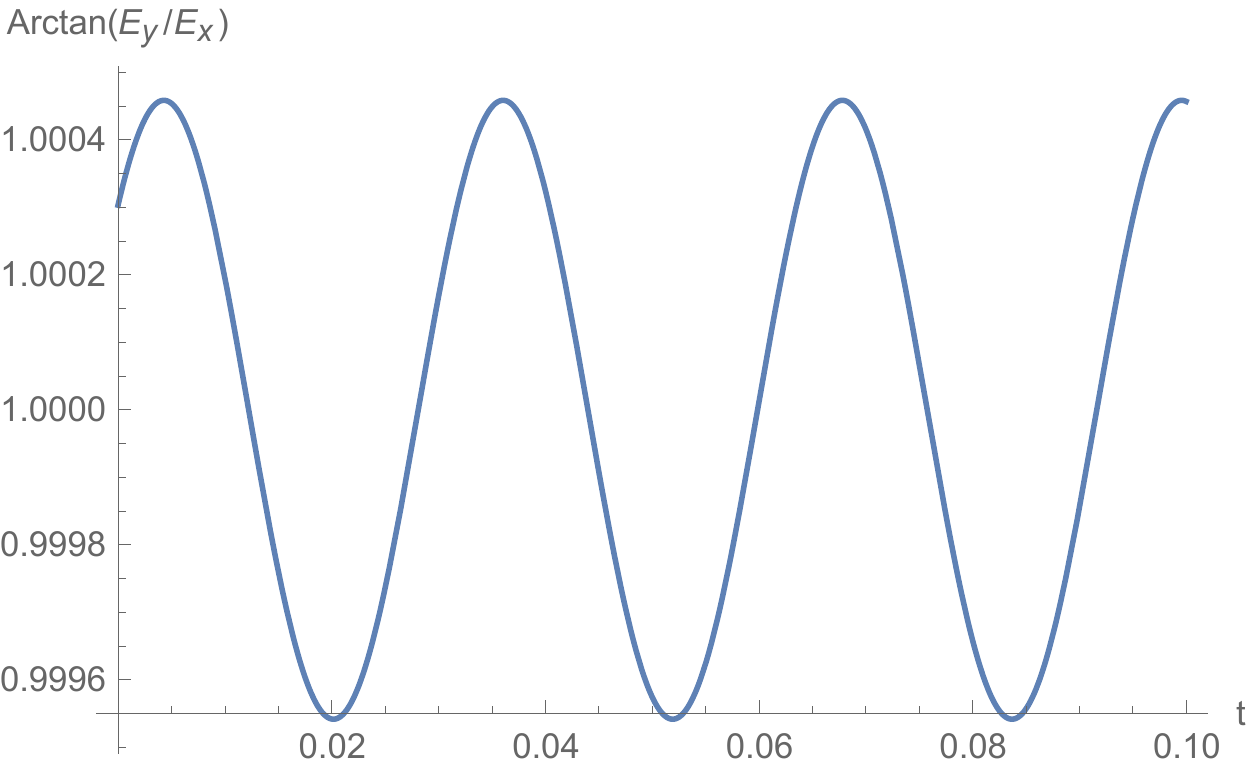}
\caption{ The dynamical polarization of the electric fluctuations generated by the GW is represented here for a fixed position in the $(x,y)$ plane. On the vertical axis we have the $\arctan (E_{y}/E_{x})$ normalized to the respective value in the absence of GW and in the horizontal axis we have time in seconds. The GW parameters are: $a=3.6\times 10^{-5}$, $b=2.6\times 10^{-5}$, $w/2\pi=31.48Hz$, $\alpha=0.26\pi$. We have used Eqs. (\ref{electricfieldxyplanex}) and (\ref{electricfieldxyplaney}), where for simplicity we assumed $\tilde{C_{1}}(z,t)=\tilde{C_{2}}(z,t)=1$ and electric field magnitudes $E_{0x}=E_{0y}=10^{-3}V/m$.}
\label{fig2}
\end{figure}

The resulting electric oscillations propagate along the $z$ axis as an electromagnetic wave with non-linear polarization. This wave results from a linear gravitational perturbation of Minkowski spacetime and therefore (in this first order approximation) it can be thought of as an electromagnetic disturbance propagating in Minkowski background with a dynamical polarization. In fact, the angle between the resulting electric field and the $x$ axis is then $\Theta\simeq \arctan \left( E_{y}/E_{x}\right) $ i.e., for $E_{0x}=E_{0y}$
\begin{eqnarray}
\Theta(x,y,z,t)\simeq \arctan \left\{ \dfrac{\left[1+\tilde{C_{2}}(z,t)e^{-\left[f_{\times}x+\left(1+f_{+}\right)y \right] }\right]}{\left[1+\tilde{C_{1}}(z,t)e^{-\left[\left(1-f_{+}\right)x+f_{\times}y \right]}\right]}\right\}.
\nonumber
\end{eqnarray}

Even if we had $\tilde{C_{1}}=\tilde{C_{2}}$, we still necessarily get a non-linear, dynamical polarization.
Such an oscillating polarization could in principle be another distinctive signature of the GW that is causing it. The solutions obtained already give sufficient information to conclude that it is possible to obtain polarization fluctuations induced by GWs, where for $E_{0x}=E_{0y}$ the strength of the effect is given by 
$\vert \pi/2-\Theta(x,y,z,t)\vert/(\pi/2)$. A dynamical spatial polarization pattern is therefore expected in our detector. This contrasts with the other cases where the resulting wave was linearly polarized. This effect is shown in Figs. \ref{fig1} and \ref{fig2}.

Nevertheless, again, the Faraday law and the Maxwell-Amp\`{e}re relations can provide constraints on the functions $\tilde{C_{1}}$ and $\tilde{C_{2}}$.

\subsubsection{Electric field in the background of a GW with zero ($\times$) mode}

If we consider solely the $(+)$ GW mode, the spacetime metric (\ref{gwmetric}) becomes diagonal and the Gauss and Maxwell-Amp\`{e}re equations simplify to the following expressions in vacuum
\begin{equation}
\partial_{k}\tilde{E}^{k}=0,\qquad \epsilon_{ijk}\partial_{j}\bar{B}^{iijjk}=\mu_{0}\jmath^{i}_{D},\label{maxampdiagonal}
\end{equation}
respectively, where
\begin{equation}
\tilde{E}^{j}\equiv -g^{jj}g^{00}\sqrt{-g}E_{j},\quad \bar{B}^{iijjk}\equiv g^{ii}g^{jj}\sqrt{-g}B^{k},
\end{equation}
and the generalized Maxwell displacement current density is
\begin{eqnarray}
\jmath^{i}_{D}
& =&\varepsilon_{0}\partial_{t}\tilde{E}^{i},\label{maxdisplacetilde}
\end{eqnarray}
in accordance with Eqs. (\ref{maxampdisplacement})-(\ref{gausschangevar2}). Let us search for a trivial electric field solution  which is fully compatible with the complete system of Maxwell equations. If we consider the field
\begin{equation}
\bold{\tilde{E}}=(\tilde{E}_{0}^{x}(y,z,t), \tilde{E}_{0}^{y}(x,z,t), \tilde{E}_{0}^{z}(x,y,t)),
\end{equation}
the Gauss law is trivially obeyed and the electric field is given by
\begin{equation}
\bold{E}=\left(\dfrac{1-f_{+}}{\sqrt{1-f_{+}^{2}}}\tilde{E}_{0}^{x},\dfrac{1+f_{+}}{\sqrt{1-f_{+}^{2}}} \tilde{E}_{0}^{y}, \dfrac{\tilde{E}_{0}^{z}}{\sqrt{1-f_{+}^{2}}}\right).\label{electricdiagonalgeneral}
\end{equation}

Furthermore, if $\partial_{t}\tilde{E}_{0}^{k}=0$ $(k=1,2,3)$, the generalized Maxwell displacement current density $\jmath^{i}_{D}$ is zero, therefore effectively the electric field does not contribute to the Maxwell-Amp\`{e}re equations. Consequently, in the absence of electric currents, such an electric field solution seems to be compatible with the condition $\bold{B}=0$. Let us assume that this is the case. Regarding the remaining Maxwell equations, the Magnetic Gauss law $\partial_{i}B^{i}=0$ is trivially obeyed but what about Faraday's law? In this case, one can show that the condition $\partial_{t}\bold{B}=-{\rm curl}\, \bold{E}=0$, leads to a field $\tilde{\bold{E}}$ which necessarily depends on time which contradicts the hypothesis of zero magnetic field according to the Maxwell-Amp\`{e}re relations in (\ref{maxampdiagonal}) and the expression (\ref{maxdisplacetilde}). In fact, one arrives at the field. 
\begin{equation}
\bold{\tilde{E}}=(\tilde{E}_{0}^{x}(z,t), \tilde{E}_{0}^{y}(z,t), \tilde{E}_{0}^{z}),
\end{equation} 
where $\tilde{E}_{0}^{z}$ is a constant and $\tilde{E}_{0}^{x}(z,t), \tilde{E}_{0}^{y}(z,t)$ are given by 
\begin{equation}
\tilde{E}_{0}^{x}=\tilde{C}_{0}^{x}\exp\left[{-\int \partial_{z}\left(\frac{1-f_{+}}{\sqrt{-g}}\right)\frac{\sqrt{-g}}{1-f_{+}}}\right],
\end{equation}
\begin{equation}
\tilde{E}_{0}^{y}=\tilde{C}_{0}^{y}\exp\left[{-\int \partial_{z}\left(\frac{1+f_{+}}{\sqrt{-g}}\right)\frac{\sqrt{-g}}{1+f_{+}}}\right],
\end{equation}
where $\tilde{C}_{0}^{x}$  and $\tilde{C}_{0}^{y}$ are constants of integration. These functions  clearly depend on time and therefore the generalized Maxwell displacement current cannot be zero leading to a non-vanishing magnetic field.
 When considering a generic electric field with three components as in (\ref{electricdiagonalgeneral}), one cannot assume that $\partial_{t}\tilde{E}_{0}^{k}=0$ neither a zero magnetic field. 

Therefore in the general case one needs to consider the influence of the electric field on the magnetic field, through the generalized Maxwell displacement current. An exception to this is the special case first considered in $\bold{III. B.1}$, where the electric field is aligned with the  direction of the propagation of the GW.

\subsection{Magnetic field oscillations induced by GWs}

The passage of the GW can induce a non-vanishing time varying magnetic field, even for an initially static electric field.
In general the full system of the Maxwell equations can be explored numerically to compute the resulting electric and magnetic oscillations. These magnetic fluctuations could be measured in principle using SQUIDS (Super Conducting Quantum Interference Devices) that are extremely sensitive to small magnetic field changes. 

To get a glimpse of the gravitationally induced magnetic field fluctuations, we can consider for simplicity only the (+) GW mode and take the generalized Maxwell-Amp\`{e}re law in the form of Eq. (\ref{maxampdisplacement}). We will be considering an electric field aligned with the $z$ axis given by the following solution to the Gauss law 
\begin{equation}
\bold{E}=\left(0,0, \dfrac{\tilde{E}_{0}^{z}(x,y,t)}{\sqrt{1-f_{+}^{2}}}\right),\qquad\partial_{k}\tilde{E}^{k}=0\,. 
\label{electriczdiag}
\end{equation}

We can also consider an electric current $I$ along the $z$ axis such that in principle, by symmetry we expect a magnetic field in the $xy$ plane, $\bold{B}=(B^{x},B^{y},0)$. Then the Maxwell-Amp\`{e}re equations (\ref{maxampdisplacement}) are
\begin{equation}
\nabla\times\bold{\bar{B}}=\mu_{0}(\sqrt{-g}\bold{j}+\varepsilon_{0}\partial_{t}\bold{\tilde{E}}_{0})\label{MaxAmpplus},
\end{equation}
where $\bold{\bar{B}}\equiv(\bar{B}^{zzyy\bold{x}},\bar{B}^{zzxx\bold{y}},0)$, while the Faraday law provides the equations
\begin{eqnarray}
\partial_{t}B^{x}=-\dfrac{\partial_{y}\tilde{E}_{0}^{z}}{\sqrt{1-f_{+}^{2}}},
\qquad 
\partial_{t}B^{y}=\dfrac{\partial_{x}\tilde{E}_{0}^{z}}{\sqrt{1-f_{+}^{2}}}.
\end{eqnarray}

Then we can perform an integration over an ``amperian'' closed line coincident to a magnetic field line (in perfect analogy with the method taken in usual electromagnetism) to integrate the Maxwell-Amp\`{e}re law, assuming axial symmetry, around the charge current distribution and electric flux (Maxwell displacement) current.

We obtain the following solution to Eq. (\ref{MaxAmpplus}) 
\begin{equation}
\bold{\bar{B}}=\dfrac{\mu_{0}\tilde{I}_{tot}}{2\pi\sqrt{x^{2}+y^{2}}}\left(\cos \phi \; \bold{e}_{y}-\sin \phi  \; \bold{e}_{x} \right), 
\end{equation}
where $\tilde{I}_{tot}(x,y,z,t)=\sqrt{-g}I+I_{D}(x,y,t)$ and $I_{D}=\int\int \jmath^{z}_{D}dxdy$. $I$ is the (constant) electric current and $\jmath^{z}_{D}=\varepsilon_{0}\partial_{t}\tilde{E}_{0}^{z}$ is the Maxwell displacement current density. We then get the magnetic field components
\begin{equation}
B^{x}=-\dfrac{1+f_{+}}{\sqrt{1-f_{+}^{2}}}\left[\dfrac{\mu_{0}\tilde{I}_{tot}(x,y,z,t)}{2\pi(x^{2}+y^{2})}y\right] ,
\end{equation}
and
\begin{equation}
B^{y}=\dfrac{1-f_{+}}{\sqrt{1-f_{+}^{2}}}\left[\dfrac{\mu_{0}\tilde{I}_{tot}(x,y,z,t)}{2\pi(x^{2}+y^{2})}x \right],
\end{equation}
respectively.
\begin{figure*}[ht]
\centering
\includegraphics[width=7.0cm]{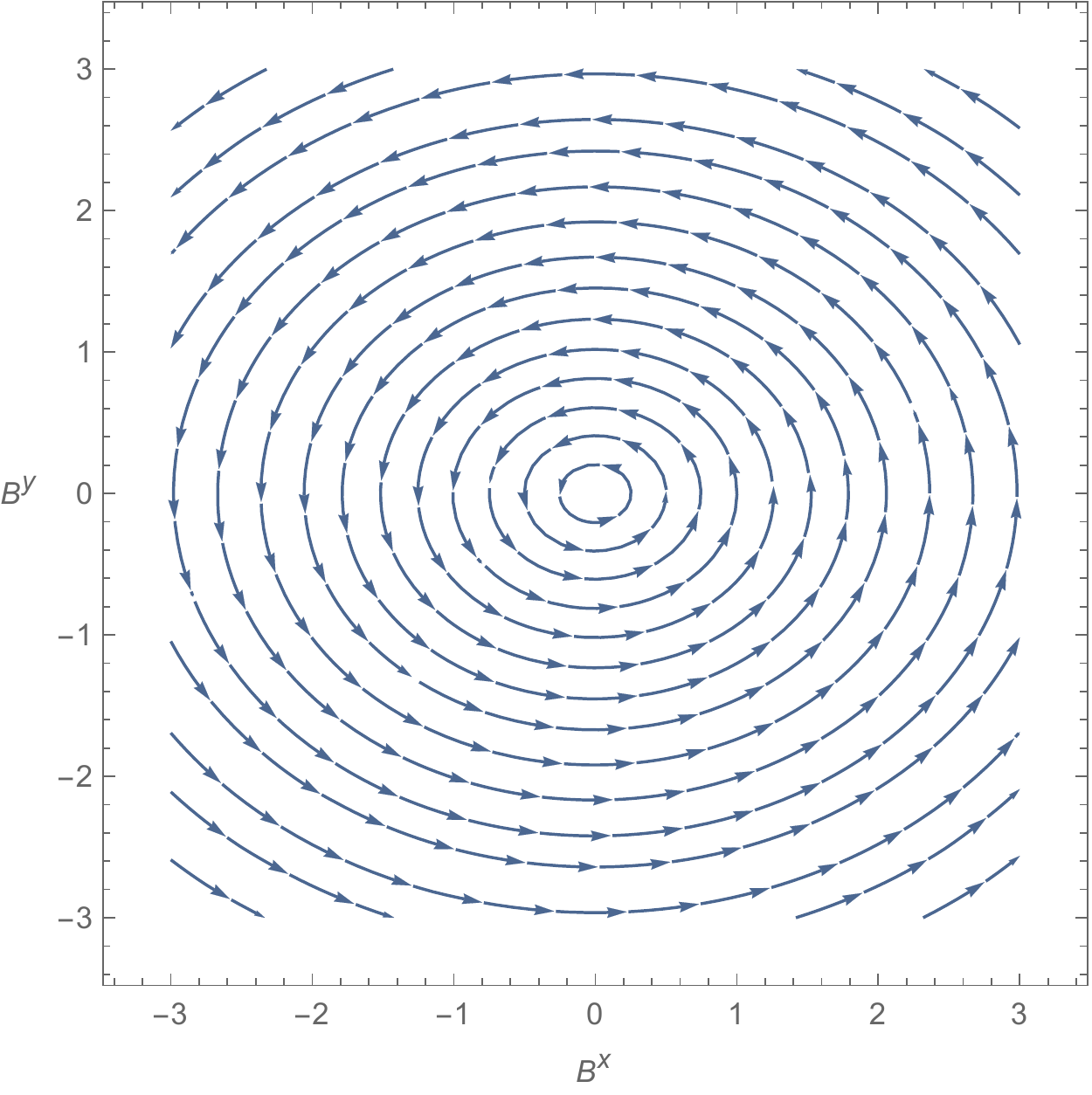} 
   \hspace{1.5cm}
\includegraphics[width=7.0cm]{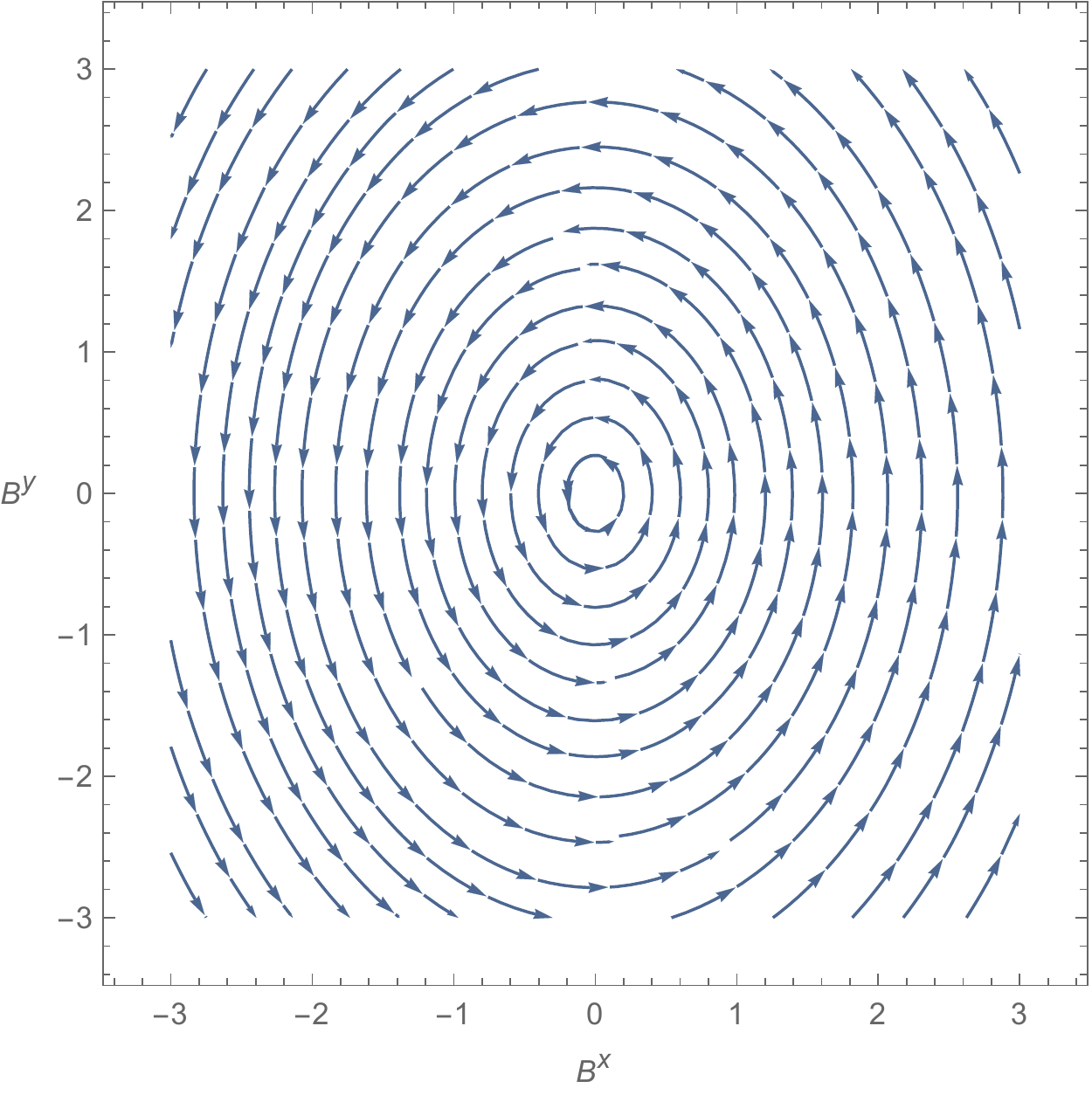} 
\caption{ These vector plots illustrate the changes of the magnetic field lines on the $(x,y)$ plane which follow the GW ($+$) mode. The two patterns are separated in time by $\tau/2$, where $\tau$ is the period. The GW parameters are: $a=0.312$, $w/2\pi=26.80Hz$. We have used expressions (\ref{gwmagx}) and (\ref{gwmagy}), with $I=4.6A$}
\label{fig3}
\end{figure*}
\begin{figure}[ht]
\centering
\includegraphics[width=7.0cm]{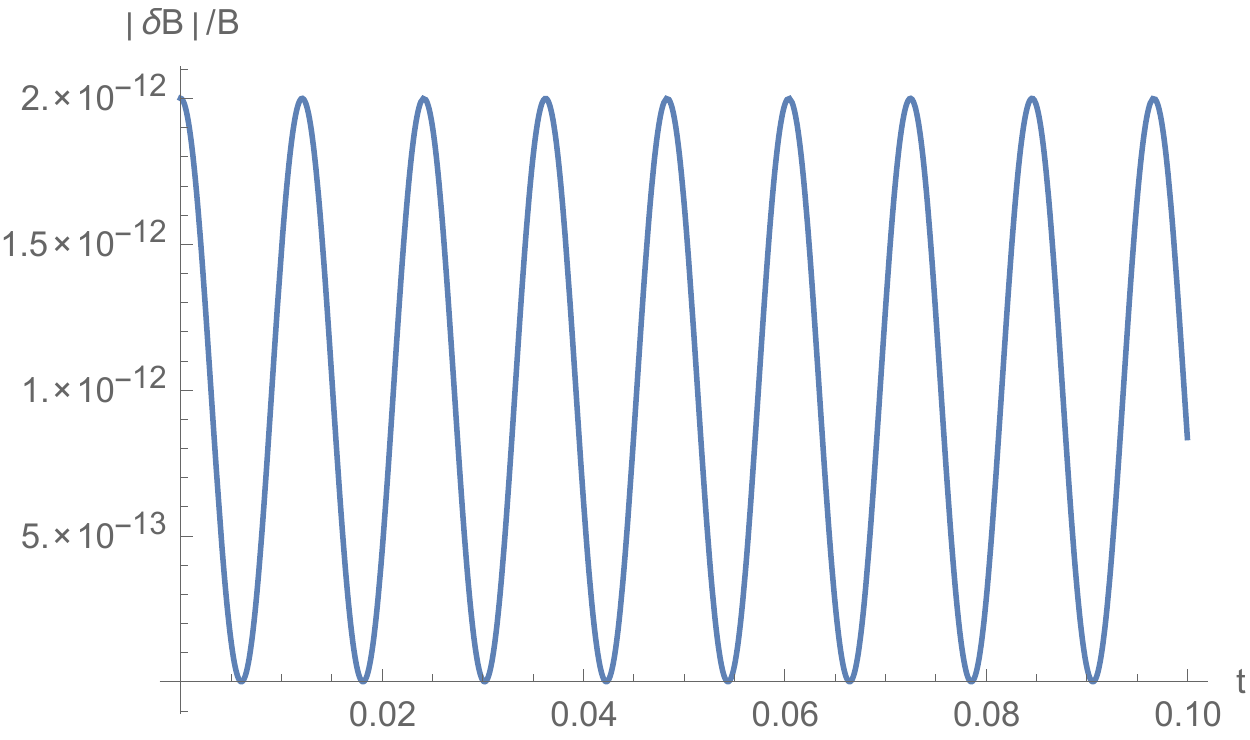}
\caption{ Here we see the strength of the magnetic fluctuation induced by the GW as a function of time (in seconds). In the vertical axis we have the non-dimensional quantity $\vert \Vert \bold{B}_{GW}\Vert - \Vert \bold{B}_{0}\Vert \vert / \Vert \bold{B}_{0}\Vert, $ where $\bold{B}_{GW}=(B^{x}, B^{y})$ is the magnetic field in the presence of the passing GW, obtained from Eqs. (\ref{gwmagx}) and (\ref{gwmagy}) and $\bold{B}_{0}$ is the control magnetic field in the case without GW. The GW parameters are: $a=2.0\times 10^{-6}$, $w/2\pi=41.38Hz$.}
\label{fig4}
\end{figure}

Making the Ansatz 
\begin{equation}
\partial_{x}\tilde{E}_{0}^{z}=-\partial_{y}\tilde{E}_{0}^{z},
\end{equation}
the Faraday equations then imply that $\partial_{t}B^{x}=\partial_{t}B^{y}$, from which one derives an equation for $\tilde{I}(x,y,z,t)$ with the general solution
\begin{eqnarray}
&&\tilde{I}= C \exp \Bigg\{ -\int 
\frac{\sqrt{1-f_{+}^{2}}}{(x-y)(1-f_{+})} \times
  \nonumber \\
&&\left[x\partial_{t}\left(\dfrac{1-f_{+}}{\sqrt{1-f_{+}^{2}}}\right) 
+y\partial_{t}\left(\dfrac{1+f_{+}}{\sqrt{1-f_{+}^{2}}}\right)\right] dt\Bigg\},
\end{eqnarray}
where $C$ is an integration constant.
In order to illustrate the general effect of the GW on the magnetic field, consider without great loss of generality that $\tilde{I}=I$ is a constant. Then the magnetic field in the background of the harmonic GW considered in (\ref{harmonicgw}) has the following fluctuations
\begin{eqnarray}
B^{x}&=&-\dfrac{\mu_{0}Iy}{2\pi(x^{2}+y^{2})}\left[1+a\cos\left(kz-wt \right)\right], \label{gwmagx}  \\
%
 B^{y}&=&\dfrac{\mu_{0}Ix}{2\pi(x^{2}+y^{2})}\left[1-a\cos\left(kz-wt \right)\right],\label{gwmagy}
\end{eqnarray}
respectively.

We can easily see that for any point $(x,y)$ fixed on any magnetic field line, the $x$ and $y$ components of the magnetic field will oscillate in time out of phase, such that when one is at its maximum value, the other is at the minimum, and vice-versa. The overall result is that the magnetic field lines will oscillate with the passage of the GW, following the deformations of the spacetime geometry, perfectly mimicking the (+) mode deformations. Figure \ref{fig3} illustrate this phenomenon and was obtained using the expressions in Eqs. (\ref{gwmagx}) and (\ref{gwmagy}). The strength of the effect as a function of time is independent from the current $I$ and it depends on the position $(x,y)$ as well as on the GW parameters (see Fig. \ref{fig4}). It can be easily shown that the strength of the fluctuations are much stronger in specific regions of the ($x,y$) plane.

\subsection{Charge density fluctuations induced by GWs}

In the previous analysis we considered the behaviour of electric and magnetic fields in vacuum regions and did not take into account the effect of the propagating GWs on charge distributions. The effect of spacetime geometry can be understood from the charge conservation equation in curved spacetime ($\nabla_{\mu}j^{\mu}=0$). As a result of this equation even in the absence of (intrinsic) currents, a non-static spacetime will induce a time variability in the charge density according to
$\partial_{t}\rho=-\partial_{t}(\log (\sqrt{-g})\rho$,
%
%
so we can write
\begin{equation}
\rho(t)=\rho_{0}\sqrt{\dfrac{g_{0}}{g(t)}},
\end{equation}
where $\rho_{0}$ is the initial charge density before the passage of the wave and $g_{0}$ is the determinant of the initially unperturbed background metric. For the simpler case of GWs travelling along the $z$ direction, seen as disturbances of Minkowski spacetime, we have the simple result
\begin{equation}
\rho(t)=\rho_{0}\left(1-f_{\times}^{2}-f_{+}^{2}\right)^{-\frac{1}{2}}.
\end{equation}
As an example, for the harmonic GW modes considered previously, we obtain 
\begin{eqnarray}
\rho(z,t)=\rho_{0}\big[1-b^{2}\cos^{2}(kz-wt+\alpha)
    \nonumber \\
-a^{2}\cos^{2}(kz-wt)\big]^{-\frac{1}{2}}.
\end{eqnarray}

Consequently, one naturally predicts charge density fluctuations and, therefore, currents due to the passage of GWs. Such density oscillations propagate along the $z$ direction following the GW penetrating a conducting material medium. This is analogous to Alfv\'{e}n waves in plasmas \cite{Vranjes:2008dz}, which are density waves induced by magnetic disturbances which propagate along the magnetic field lines. In this case, astrophysical sources of GWs such as Gamma Ray Bursts or generic coalescing binaries that happen to be surrounded by plasmas in accretion disks or in stellar atmospheres, might generate similar mass density waves and charge density waves induced by the GW propagation.  A more realistic treatment would require the equations of Magneto-Hydrodynamics in the background of a GW (see \citep{Brodin:2000du}). An interesting study would be to consider the backreaction of the relativistic plasma and electromagnetic fields on the GW properties such as the frequency, amplitude and polarizations, so that the travelling wave after detection could, in principle, contain information about the physical properties of the medium through which it propagated.

The above expression can also indicate another window for GW detection. Conductors in  perfect electrostatic equilibrium or superconducting materials at very low temperatures might reveal very dim electric oscillations with well-defined characteristics, induced by GWs.

\subsection{GW effects on electromagnetic radiation}

The vacuum equations for the 4-potential in the presence of a background GW can be derived from Eq. (\ref{Proca}). In terms of the electric and magnetic components of the 4-potential, we have 
\begin{eqnarray}
\partial_{\mu}\partial^{\mu}\phi+\dfrac{f_{\times}\left(\partial_{t}f_{\times} \right)+f_{+}\left(\partial_{t}f_{+} \right) }{f_{\times}^{2}+f_{+}^{2}-1}
\partial_{t}\phi
    \nonumber  \\
-\dfrac{f_{\times}\left(\partial_{z}f_{\times} \right)+f_{+}\left(\partial_{z}f_{+} \right) }{f_{\times}^{2}+f_{+}^{2}-1}
\partial_{z}\phi=0,
\end{eqnarray}
and
\begin{eqnarray}
\partial_{\mu}\partial^{\mu}A^{k}+\dfrac{f_{\times}\left(\partial_{t}f_{\times} \right)+f_{+}\left(\partial_{t}f_{+} \right) }{f_{\times}^{2}+f_{+}^{2}-1}
\partial_{t}A^{k}
   \nonumber  \\
-\dfrac{f_{\times}\left(\partial_{z}f_{\times} \right)+f_{+}\left(\partial_{z}f_{+} \right) }{f_{\times}^{2}+f_{+}^{2}-1}\partial_{z}A^{k}=0,
\end{eqnarray}
respectively. In the absence of GWs we recover the usual wave equations.
 
The resulting expressions simplify significantly if one considers only one of the two possible GW modes. For example, for an electromagnetic wave travelling in the $z$ direction and the harmonic GW in Eq. (\ref{harmonicgwa}) with no ($\times$) mode, we get the following wave equation for the electric potential
\begin{eqnarray}
\partial^{2}_{tt}\phi &-&\partial^{2}_{zz}\phi-
\dfrac{wa^{2}\sin(wt-kz)\cos(wt-kz) }{a^{2}\cos^{2}(wt-kz)-1}\partial_{t}\phi
    \nonumber \\
&-&\dfrac{ka^{2}\sin(wt-kz)\cos(wt-kz) }{a^{2}\cos^{2}(wt-kz)-1}
\partial_{z}\phi=0,\label{gwpotentialwave}
\end{eqnarray}
that can be studied by applying Fourier transformation methods. 

In order to study in depth the physical (measurable) effects of the passage of the GW on electromagnetic wave dynamics, one needs to solve these equations and then compute the gauge invariant electric and magnetic fields. We see in the above wave equations, the presence of terms proportional to the first derivatives which are completely absent in the electromagnetic wave equations in flat spacetime (in cartesian coordinates). These terms are always induced by gravitational fields, but in this case the gravitational field is dynamical which represents a much richer electromagnetic wave signal with the signature of the GW (see also \cite{Marklund:1999sp}). Such signals in the radio regime might possibly be detectable through methods of Long Baseline Interferometry, in order to amplify it. Nevertheless, we can see from the expressions above that the extra terms on the electromagnetic wave equations, induced by GWs are proportional to the frequency. Such gravitational effects might become important for sufficiently hight frequency GWs. Simulations are required to see the feasibility or not of such methods.  

\section{Discussion and Final remarks}\label{secIV}

GW astronomy is an emerging field of science with the potential to revolutionize astrophysics and cosmology. The construction of GW observatories can also effectively boost major technological developments.  Given the extremely low GW amplitudes reaching the Solar System, incredibly huge laser interferometers have been built and others are under development in order to reach the required sensitivities. In fact, the biggest of all, LISA is expected to be achieved in space possibly through an ESA-NASA collaboration. ESA's LISA-Pathfinder science mission officially started on March 2016 and during the following six months it conducted many experiments to pave the way for future space GW observatories, such as LISA.  These huge observatories represent an amazing technological effort. A natural question is the following: Can we amplify the GW signal?

One fundamental prediction of the coupling between gravity and electromagnetism is the generation of electromagnetic waves due to gravitational radiation. Therefore, in principle under the appropriate resonant conditions, the electromagnetic signal thus produced can be amplified allowing us to measure GWs, not through the motion of test masses but rather by transferring the GW signal directly into electromagnetic information. This fact might represent an important change in perspective for future ground and space GW detectors.

The fact that GWs can generate electromagnetic waves is of course not evident if one restricts the analysis to the propagation of light rays (in the geometrical optics limit) in curved spacetime. On the other hand, the full Einstein-Maxwell system of equations have to take into account the curved spacetime within Maxwell's equations and also the contribution of the electromagnetic stress-energy tensor to the gravitational field. The first aspect of this coupling was considered in this work, and it is sufficient to show that GWs can be sources of electromagnetic waves. The full gravity-electromagnetic coupling also shows the reverse phenomenon.

In this work, we obtained electric and magnetic field oscillations fully induced by a GW travelling along the $z$ axis. For simplicity we assumed harmonic GWs.
We considered the Gauss law for the cases of an electric field along the $z$ axis, along the $x$ axis and in the $(x,y)$ plane. In the first case, the solutions in Eqs. (\ref{gwelectricz}) and (\ref{gwelectricz2}) allowed to make an estimation of the energy flux of the resulting radiation. It is important to emphasize the fact that the electric fluctuation thus produced corresponds to a longitudinal wave. This means that a non-zero longitudinal mode in electromagnetic radiation can in general be induced by gravitational radiation. One should search for these GW signatures in the electromagnetic counterpart of GW sources. The solution we obtained shows the dependence on the amplitudes of the two GW modes, $a$ and $b$, as well as on the frequency $w$ and phase difference $\alpha$. 
An important aspect of hypothetical electric-GW detectors is the fact that in general although the signal is very weak for any GW reaching the solar system, under appropriate resonant conditions it can be amplified. In fact, this can be used to improve the signal to noise ratio since a system analogous to optical resonators can act as a filter privileging the signal with a specific (resonant) frequency.
  
The changing electric field in Eq. (\ref{gwelectricz}) inside a capacitor, for instance, would also generate alternate currents in any conductor placed between the capacitor's charged plates. In particular, a diode placed in the appropriate orientation would allow a current signal in a single direction intermittently, following the rhythm of the GW fluctuations.
In the ($x,y$) plane the electric field can be generated by two independent capacitors in perpendicular configuration. The approximate solutions obtained show electric field oscillations generated by the GW which propagate along the $z$ axis with non-linear polarization. We can expect a spatial polarization pattern in our detector which changes with time. This contrasts with the other cases where the resulting wave was linearly polarized. 
This effect is shown in Figs. \ref{fig1} and \ref{fig2}.

In all cases, the resulting electromagnetic signal has the signature of the GW that produces it, depending on $a$, $b$, $w$ and $\alpha$. In any of these examples, time varying electric fields are generated, which can contribute to the magnetic field via the Maxwell-Amp\`{e}re law. In particular, they appear in the generalized Maxwell displacement current vector density, Eq. (\ref{maxdisplacement}), induced by the GW. This in turn can generate a time varying magnetic field even in the absence of electric currents. Accordingly, GWs also induce magnetic field oscillations. 
We made an estimation of such an effect considering the case of a diagonal metric by setting the ($\times$) GW mode to zero. We assumed a certain electric current $I$ along the $z$ axis and the electric field in Eq. (\ref{electriczdiag}) along the same direction. The magnetic field thus generated lies on the $(x,y)$ plane and it is easy to see that for any point $(x,y)$ fixed on the magnetic field lines, the $x$ and $y$ components of the magnetic field will oscillate in time out of phase, such that when one is at its maximum value, the other is at the minimum, and vice-versa. 

The overall result is that the magnetic field lines will oscillate with the passage of the GW, following the deformations of spacetime geometry. Figure 3 illustrate this phenomenon and was obtained using the expressions in Eqs. (\ref{gwmagx}) and (\ref{gwmagy}).  In Fig. 4 we see the strength of the effect as a function of time. The signal to be measured is independent from the current $I$ and it depends on the position $(x,y)$ as well as on the GW parameters. It can be easily shown that the strength of the fluctuations are much stronger in specific regions of the ($x,y$) plane. Such small magnetic field changes could in principle be measured with SQUIDS (SuperConducting Quantum Interference Devices), which are sensitive to extremely small magnetic fields changes \cite{squids1}-\cite{Granata:2015qga}.  SQUIDS have amazing applications form biophysics (in particular to biomagnetism) and medical sciences but also to theoretical physics: studies of majorana fermions \cite{Rahmonov:2016gfu}, dark matter \cite{Popov:2014mba}, gravity wave resonant bar detectors \cite{squidsgws}, cosmological fluctuations \cite{Beck:2014sia,Beck:2016npj}. 

The calculations in this work point to electromagnetic effects induced by GWs such that
\begin{equation}
\dfrac{\delta \bold{E}}{\bold{E}}\sim h, \qquad \dfrac{\delta \bold{B}}{\bold{B}}\sim h, \qquad h\sim 10^{-21},
\end{equation}
where $h$ is the amplitude (strain) of the GW reaching the Solar System. SQUIDS have an incredible sensitivity \cite{squids1}-\cite{Takeuchi:2013kra} being able to measure magnetic fields of the order of $10^{-15} T$ or even $10^{-18} T$ for measurements performed over a sufficient period of time (the SQUIDS used in the GP-B experiment had this sensitivity). Using these values for the SQUIDS sensitivity, in order to be able to measure the tiny GW effects on magnetic fields we would require magnetic fields of the order of $\bold{B}\sim10^{6} T$ or in the best case $\bold{B}\sim10^{3} T$. Presently, the highest magnetic fields produced in the laboratory have values of $\bold{B}\sim 45 T$ (continuous) and $\bold{B}\sim 100 T-10^{3} T$ (pulsed). therefore, although SQUIDS are extremely sensitive, there is a real limitation to perform these measurements coming from the huge magnetic fields required. Nevertheless, the science of SQUIDS and ultra-sensitive magnetometers is very active and evolving \cite{Storm:2017muk,Takeuchi:2013kra} and it is natural to expect improvements in terms of sensitivities and noise reduction and modelling. For $\bold{B}\sim 10 T$ laboratory magnetic fields we would require extremely higher sensitivities ($\delta \bold{B}\sim 10^{-20}$) which is not in the reach of present magnetometers. Besides these considerations, intrinsic and extrinsic noise should be extremely well modelled and if possible reduced by advanced cryogenics and filtering processes.

We may consider the use of electromagnetic cavity resonators to amplify the electromagnetic waves induced by the GWs. For magnetometers  with $\delta \bold{B}\sim 10^{-18}T$ sensitivities and $10 T$ reference magnetic fields, it means that the amplification of the signal would have to be about 2 orders of magnitude. Even if this cannot be achieved by present day electromagnetic resonators it might be in the near future. An important advantage of these cavities is that in practice they work has filters being able to amplify a signal centred around a specific frequency which corresponds to the fundamental frequency of the resonator. For cylindrical resonators with size $L$, the wavelength of the fundamental frequency is $\lambda\sim 2L$, meaning that different resonators of different sizes would be sensitive to the different parts of the GW spectrum. By effectively filtering and amplifying the signal around a certain frequency far from the noise peak, it is in principle possible to increase substantially the signal to noise ratio, which is essential for a good measurement/detection.

Let us consider the case where we se electric fields instead of magnetic fields in our electromagnetic detectors, for example the electric field inside a charged plane capacitor. By measuring the Voltage signal instead of electric field, we have the advantage of being able with the present technology to, on one hand, easily produce $10^{3}V$ or higher static fields and on the other hand, to reach sensitivities of $\delta V\sim 10^{-15}V$. This means that the signal should be amplified 2 to 3 orders of magnitude. The combination of electromagnetic cavity resonators and electronic amplifiers (for the Voltage signal) could make this a real possibility for GW detectors.

Moving now from human made laboratories on earth or in space to natural astrophysical observatories, we call the attention to the fact that the highest magnetic field values (indirectly) measured so far are those of neutron stars with values around $10^{6}T - 10^{11}T$. Radio and X-ray astronomy is able to indirectly measure these astrophysical magnetic fields by considering the propeties of Cyclotron radiation. A stochastic GW background signal due to inumerable sources in the galaxy and beyond is expected to leave a measurable imprint on the magnetic field of normal pulsars and magnetars. In fact, this method could be used in a complementary way to that of PTA (Pulsar Timing Arrays) to measure a stochastic GW signal. The huge magnetic fields in the surroundings of pulsars makes them natural laboratories to study the effects of GWs on electromagnetic fields. The use of arrays of Pulsars could be advantageous in order to distinguish the GW signal from intrinsic fluctuations of the magnetic field and to better deal with extrinsic noise. Pulsars are extremely precise clocks and if they behave as very stable dynamos, then it might be possible to generalize the methods and years of improvement in PTA by measuring the interaction of GWs with magnetic fiels. Is this another window for GW astronomy through VLBL (Very Long Baseline Interferometry)? We leave this as an open question that deserves more research from both theorists and observation experts.

We also obtained charge density oscillations induced by GWs. These can propagate as density waves in the case of charged fluids, through which a GW is propagating. This effect deserves to be taken in consideration within more complete magnetohydrodynamical computations, in order to have  simulations of the effects of GWs in plasmas near the cores of highly energetic GW sources. These plasma environments might occur in different astrophysical sources such as Gamma Ray Bursts and some specific coalescing binaries.

Regarding electromagnetic waves in the presence of gravity, extra terms appear in the generalized wave equations which deserves further research to get a full analysis of the approximate solutions. Indeed, going beyond the geometrical optics limit, light deflection (null geodesics) and gravitational redshift are not the only effects arising from the coupling between light and gravity. More generally, all electromagnetic waves can experience gravitational effects on the amplitudes, frequencies and polarizations. Besides, as shown in \cite{FCFLapplications}, electric and magnetic wave dynamics can be coupled due to the non-stationary geometries, as is the case of GWs. Important studies have been made regarding the electromagnetic counterpart of GW sources (see for example \cite{Capozziello:2010sm} and \cite{Capozziello:2011fs}), but there is much to explore in the landscape of (multi-messenger) gravitational and electromagnetic astronomy  

In general, one expects that GWs induce very rich  electromagnetic wave dynamics. These effects might become more significant for very high frequency GWs as one can see from Eq. (\ref{gwpotentialwave}). Moreover, the terms proportional to the first derivatives of the 4-potential have space and time varying coefficients. For the harmonic GWs considered in this work, these coefficients oscillate between positive and negative numbers, a fact that might imply a very distinctive wave modulation pattern of the resulting electromagnetic wave. This hypothesis and its implications requires further investigation as it might provide very rich GW information codified in the electromagnetic spectra of different astrophysical and even cosmological sources.

\section*{Acknowledgments}
FC acknowledges financial support of the Funda\c{c}\~{a}o para a Ci\^{e}ncia (FCT) through the grant PD/BD/128017/2016 and Programa de Doutoramento FCT, PhD::SPACE – Doctoral Network for Space Sciences (PD/00040/2012).
FSNL acknowledges financial support of an Investigador FCT Research contract, with reference IF/00859/2012, funded by FCT/MCTES (Portugal).
This article is based upon work from COST Action CA15117, supported by COST (European Cooperation in Science and Technology).




\begin{thebibliography}{10}

\bibitem{Abbott:2016blz}
  B.~P.~Abbott {\it et al.} [LIGO Scientific and Virgo Collaborations],
  ``Observation of Gravitational Waves from a Binary Black Hole Merger,''
  Phys.\ Rev.\ Lett.\  {\bf 116} (2016),  061102
  [arXiv:1602.03837 [gr-qc]].

\bibitem{e-LISA}
  P.~Amaro-Seoane {\it et al.},
  ``eLISA/NGO: Astrophysics and cosmology in the gravitational-wave millihertz regime,''
  GW Notes {\bf 6} (2013) 4
  [arXiv:1201.3621 [astro-ph.CO]].

\bibitem{Colpi:2016fup}
  M.~Colpi and A.~Sesana,
  ``Gravitational wave sources in the era of multi-frequency gravitational wave astronomy,''
  arXiv:1610.05309 [astro-ph.HE].


\bibitem{Sathyaprakash:2009xs}
  B.~S.~Sathyaprakash and B.~F.~Schutz,
  ``Physics, Astrophysics and Cosmology with Gravitational Waves,''
  Living Rev.\ Rel.\  {\bf 12} (2009) 2
  [arXiv:0903.0338 [gr-qc]].

\bibitem{Bogdanos:2009tn}
  C.~Bogdanos, S.~Capozziello, M.~De Laurentis and S.~Nesseris,
  ``Massive, massless and ghost modes of gravitational waves from higher-order gravity,''
  Astropart.\ Phys.\  {\bf 34} (2010) 236
  [arXiv:0911.3094 [gr-qc]].

\bibitem{Cardoso:2016oxy}
  V.~Cardoso, S.~Hopper, C.~F.~B.~Macedo, C.~Palenzuela and P.~Pani,
  ``Gravitational-wave signatures of exotic compact objects and of quantum corrections at the horizon scale,''
  Phys.\ Rev.\ D {\bf 94} (2016) no.8,  084031
  [arXiv:1608.08637 [gr-qc]].
  
\bibitem{Abedi:2016hgu}
  J.~Abedi, H.~Dykaar and N.~Afshordi,
  ``Echoes from the Abyss: Evidence for Planck-scale structure at black hole horizons,''
  arXiv:1612.00266 [gr-qc].
  
\bibitem{Zhu:2015ara}
  X.~J.~Zhu, L.~Wen, G.~Hobbs, R.~N.~Manchester and R.~M.~Shannon,
  ``Detecting nanohertz gravitational waves with pulsar timing arrays,''
  arXiv:1509.06438 [astro-ph.IM].

\bibitem{Barausse:2014tra}
  E.~Barausse, V.~Cardoso and P.~Pani,
  ``Can environmental effects spoil precision gravitational-wave astrophysics?,''
  Phys.\ Rev.\ D {\bf 89} (2014),  104059
  [arXiv:1404.7149 [gr-qc]].

\bibitem{GWsRiles:2012yw}
  K.~Riles,
  ``Gravitational Waves: Sources, Detectors and Searches,''
  Prog.\ Part.\ Nucl.\ Phys.\  {\bf 68} (2013) 1
  [arXiv:1209.0667 [hep-ex]].

\bibitem{Singer:2014qca}
  L.~P.~Singer {\it et al.},
  ``The First Two Years of Electromagnetic Follow-Up with Advanced LIGO and Virgo,''
  Astrophys.\ J.\  {\bf 795} (2014) 2,  105
  [arXiv:1404.5623 [astro-ph.HE]].

\bibitem{Hacyan:2015kra}
S.~Hacyan,
  ``Effects of gravitational waves on the polarization of pulsars,''
  Int.\ J.\ Mod.\ Phys.\ A {\bf 31}, no. 02n03, 1641023 (2016)
  [arXiv:1502.04630 [gr-qc]].

\bibitem{Brodin:2000du}
  G.~Brodin, M.~Marklund and P.~K.~S.~Dunsby,
  ``Nonlinear gravitational wave interactions with plasmas,''
  Phys.\ Rev.\ D {\bf 62} (2000) 104008
  [gr-qc/0006030].

\bibitem{GravitomagnetismMashhoon:2003ax}
  B.~Mashhoon,
  ``Gravitoelectromagnetism: A Brief review,''
  gr-qc/0311030.
  
\bibitem{Everitt:2011hp}
  C.~W.~F.~Everitt {\it et al.},
  ``Gravity Probe B: Final Results of a Space Experiment to Test General Relativity,''
  Phys.\ Rev.\ Lett.\  {\bf 106} (2011) 221101
  [arXiv:1105.3456 [gr-qc]].
  
\bibitem{Tajmar:2004ww}
  M.~Tajmar and C.~J.~de Matos,
  ``Extended analysis of gravitomagnetic fields in rotating superconductors and superfluids,''
  Physica C {\bf 420} (2005) 56
  [gr-qc/0406006].

\bibitem{Tajmar:2006gh}
  M.~Tajmar, F.~Plesescu, B.~Seifert and K.~Marhold,
  ``Measurement of Gravitomagnetic and Acceleration Fields Around Rotating Superconductors,''
  AIP Conf.\ Proc.\  {\bf 880} (2007) 1071
  [gr-qc/0610015].

\bibitem{Marklund:1999sp}
  M.~Marklund, G.~Brodin and P.~K.~S.~Dunsby,
  ``Radio wave emissions due to gravitational radiation,''
  Astrophys.\ J.\  {\bf 536} (2000) 875
  [astro-ph/9907350].


\bibitem{Montanari:1998gd}
  E.~Montanari,
  ``On the propagation of electromagnetic radiation in the field of a plane gravitational wave,''
  Class.\ Quant.\ Grav.\  {\bf 15} (1998) 2493
  [gr-qc/9806054].
  
\bibitem{Rakhmanov:2009zz}
  M.~Rakhmanov,
  ``On the round-trip time for a photon propagating in the field of a plane gravitational wave,''
  Class.\ Quant.\ Grav.\  {\bf 26} (2009) 155010
  [arXiv:1407.5376 [gr-qc]].


\bibitem{FCFL}
F.~Cabral and F.~S.~N.~Lobo,
  ``Electrodynamics and spacetime geometry I: Foundations,''
  Found.\ Phys.\  {\bf 47}, no. 2, 208 (2017)
  [arXiv:1602.01492 [gr-qc]].
  
  
\bibitem{FCFLapplications}
  F.~Cabral and F.~S.~N.~Lobo,
  ``Electrodynamics and spacetime geometry: Astrophysical applications,''
  arXiv:1603.08180 [gr-qc].

\bibitem{HehlYuribook}
F.~W.~Hehl and Y.~N.~Obukhov,
``Foundations of Classical Electrodynamics: Charge, Flux, and Metric'',
 Birkhäuser, Boston, 2003.

\bibitem{Gronwald:2005tv}
  F.~Gronwald, F.~W.~Hehl and J.~Nitsch,
  ``Axiomatics of classical electrodynamics and its relation to gauge field theory,''
  physics/0506219.

\bibitem{Hehl:2000pe}
  F.~W.~Hehl and Y.~N.~Obukhov,
  ``A gentle introduction to the foundations of classical electrodynamics: The meaning of the excitations (D,H) and the field strengths (E, B),''
  physics/0005084.

\bibitem{Hehl:2005hu}
  F.~W.~Hehl and Y.~N.~Obukhov,
  ``Spacetime metric from local and linear electrodynamics: A New axiomatic scheme,''
  Lect.\ Notes Phys.\  {\bf 702} (2006) 163
  [gr-qc/0508024].

\bibitem{Hehl:1999bt}
  ``F.~W.~Hehl and Y.~N.~Obukhov,
  How does the electromagnetic field couple to gravity, in particular to metric, nonmetricity, torsion, and curvature?'',
  Lect.\ Notes Phys.\ {\bf 562} (2001) 479, [gr-qc/0001010].
  

\bibitem{Rezzolla:2000dk}
  L.~Rezzolla, B.~J.~Ahmedov and J.~C.~Miller,
  ``General relativistic electromagnetic fields of a slowly rotating magnetized neutron star,''
  Mon.\ Not.\ Roy.\ Astron.\ Soc.\  {\bf 322} (2001) 723
  [astro-ph/0011316].
 
\bibitem{Vranjes:2008dz} 
  J.~Vranjes, S.~Poedts, B.~P.~Pandey and B.~De Pontieu,
  ``Energy flux of Alfven waves in weakly ionized plasma,''
  Astron.\ Astrophys.\  {\bf 478}, 553 (2008)
  [arXiv:0805.4591 [astro-ph]]. 
 

\bibitem{squids1}  
A.~Barone and G.~Patern\`{o}, ``Physics and Applications of the Josephson Effect'', Wiley, New York, 1982.

\bibitem{squids2}  
M.~Tinkham, ``Introduction to Superconductivity'', Dover, New York, 1996.

\bibitem{squids3} 
J.~Clarke,
  ``SQUIDs: then and now,''
  Int.\ J.\ Mod.\ Phys.\ B {\bf 24}, 3999 (2010).
 

\bibitem{Granata:2015qga}
  C.~Granata and A.~Vettoliere,
  ``Nano Superconducting Quantum Interference device: a powerful tool for nanoscale investigations,''
  Phys.\ Rept.\  {\bf 614} (2016) 1 
  [arXiv:1505.06887 [cond-mat.supr-con]].


\bibitem{Rahmonov:2016gfu}
  I.~R.~Rahmonov, Y.~M.~Shukrinov, R.~Dawood and H.~E.~Samman,
  ``Determination of Cooper pairs and Majorana fermions currents ratio in DC-SQUID with topologically nontrivial barriers,''
  arXiv:1611.07179 [cond-mat.supr-con].
  
\bibitem{Popov:2014mba}
  V.~Popov,
  ``Resonance detection of dark matter axions using a DC SQUID,''
  J.\ Exp.\ Theor.\ Phys.\  {\bf 122} (2016) no.2,  236
   [Zh.\ Eksp.\ Teor.\ Fiz.\  {\bf 149} (2016) no.2,  272]
  [arXiv:1410.6682 [hep-ph]].    


\bibitem{squidsgws}
  A.~Vinante,
  ``Optimization of a Two-Stage dc SQUID for Resonant Gravitational Wave Detectors,''  PhD thesis (2016).
  
  
\bibitem{Beck:2016npj}
  C.~Beck,
  ``Cosmological flux noise and measured noise power spectra in SQUIDs,''
  Sci.\ Rep.\  {\bf 6} (2016) 28275.
  
  
\bibitem{Beck:2014sia}
  C.~Beck,
  ``Magnetic flux noise in q-bits and SQUIDS from primordial density fluctuations,''
  arXiv:1409.4759 [physics.gen-ph].
  

 
\bibitem{Storm:2017muk}
  J.~H.~Storm, P.~Hömmen, D.~Drung and R.~Körber,
  ``An ultra-sensitive and wideband magnetometer based on a superconducting quantum interference device,''
  Appl.\ Phys.\ Lett.\  {\bf 110} (2017) 072603
  [arXiv:1702.05428 [physics.ins-det]].  
  
  
\bibitem{Takeuchi:2013kra}
  S.~Takeuchi,
  ``Holographic Superconducting Quantum Interference Device,''
  Int.\ J.\ Mod.\ Phys.\ A {\bf 30} (2015) no.09,  1550040
  [arXiv:1309.5641 [hep-th]].


\bibitem{Capozziello:2010sm}
  S.~Capozziello, M.~De Laurentis, I.~De Martino and M.~Formisano,
  ``Short Gamma Ray Bursts as possible electromagnetic counterpart of coalescing binary systems,''
  Astrophys.\ Space Sci.\  {\bf 332} (2011) 31
  [arXiv:1004.4818 [astro-ph.CO]].  


\bibitem{Capozziello:2011fs}
  S.~Capozziello, M.~De Laurentis, I.~De Martino, M.~Formisano and D.~Vernieri,
  ``Gravitational and electromagnetic emission by magnetized coalescing binary systems,''
  Astrophys.\ Space Sci.\  {\bf 333} (2011) 29
  [arXiv:1101.5306 [astro-ph.HE]].

\end{thebibliography}
\end{document}